\documentclass{emulateapj}

\usepackage{float}
\usepackage{amsmath}
\usepackage{epsfig,floatflt}
\usepackage{subfigure}

\newcommand{\npix}{N_{\textrm{pix}}}
\newcommand{\BA}{\mathbf{A}}
\newcommand{\BB}{\mathbf{B}}
\newcommand{\BC}{\mathbf{C}}
\newcommand{\Ba}{\mathbf{a}}
\newcommand{\Bb}{\mathbf{b}}
\newcommand{\Bc}{\mathbf{c}}
\newcommand{\Bs}{\mathbf{s}}
\newcommand{\BS}{\mathbf{S}}
\newcommand{\Bn}{\mathbf{n}}
\newcommand{\BN}{\mathbf{N}}

\newcommand{\BF}{\mathbf{F}}
\newcommand{\Bd}{\mathbf{d}}

\newcommand{\Bx}{\mathbf{x}}

\newcommand{\Bu}{\mathbf{u}}

\newcommand{\BD}{\mathbf{D}}

\newcommand{\BT}{\mathbf{T}}
\newcommand{\BG}{\mathbf{G}}

\newcommand{\Cell}{C_{\ell}}
\newcommand{\muK}{\mu\textrm{K}}

    \newcommand{\qed}{\nobreak \ifvmode \relax \else
          \ifdim\lastskip<1.5em \hskip-\lastskip
          \hskip1.5em plus0em minus0.5em \fi \nobreak
          \vrule height0.75em width0.5em depth0.25em\fi}

\begin{document}
\title{Joint Bayesian component separation and CMB power spectrum estimation}


\author{H.\ K.\ Eriksen\altaffilmark{1,2,3}, J. B.
  Jewell\altaffilmark{4}, C. Dickinson\altaffilmark{4},
  A. J. Banday\altaffilmark{5}, K. M. G\'{o}rski\altaffilmark{4,6},
  C. R. Lawrence\altaffilmark{4}}

\altaffiltext{1}{email: h.k.k.eriksen@astro.uio.no}

\altaffiltext{2}{Institute of Theoretical Astrophysics, University of
Oslo, P.O.\ Box 1029 Blindern, N-0315 Oslo, Norway}

\altaffiltext{3}{Centre of
Mathematics for Applications, University of Oslo, P.O.\ Box 1053
Blindern, N-0316 Oslo}

\altaffiltext{4}{Jet Propulsion Laboratory, California Institute
of Technology, Pasadena CA 91109} 

\altaffiltext{5}{Max-Planck-Institut f\"ur Astrophysik,
Karl-Schwarzschild-Str.\ 1, Postfach 1317, D-85741 Garching bei
M\"unchen, Germany}

\altaffiltext{6}{Warsaw University Observatory, Aleje Ujazdowskie 4, 00-478 Warszawa,
  Poland}

\date{Received - / Accepted -}

\begin{abstract}
  We describe and implement an exact, flexible, and computationally
  efficient algorithm for joint component separation and CMB power
  spectrum estimation, building on a 
  Gibbs sampling framework. Two essential new features are 1)~conditional 
  sampling of foreground spectral parameters, and 2)~joint
  sampling of all amplitude-type degrees of freedom (e.g., CMB,
  foreground pixel amplitudes, and global template amplitudes) given
  spectral parameters. Given a parametric model of the foreground
  signals, we estimate efficiently and accurately the exact joint
  foreground-CMB posterior distribution, and therefore all marginal distributions 
  such as the
  CMB power spectrum or foreground spectral index posteriors. The main
  limitation of the current implementation is the requirement of
  identical beam responses at all frequencies, which restricts the
  analysis to the lowest resolution of a given experiment.  We outline
 a future generalization to multi-resolution observations. 
 To verify the method,  we analyse simple models and compare the results to
  analytical predictions. We then analyze a realistic simulation with
  properties similar to the 3-yr WMAP data, downgraded to a common
  resolution of $3^{\circ}$ \hbox{FWHM}.  The results from the actual 3-yr
  WMAP temperature analysis are presented in a companion Letter.
\end{abstract}


\maketitle

\section{Introduction}

Great advances have been made recently both in experimental techniques
for studying the cosmic microwave background (CMB) and in the
measurements themselves.  The angular power spectrum of temperature
fluctuations has been characterized over more than three decades in
angular scale \citep{hinshaw:2007, kuo:2007, readhead:2004}, and even
the E-mode polarization spectrum has now been measured to some
precision \citep{ade:2007, page:2007, montroy:2006, sievers:2007}.  In
the coming years, even greater improvements in sensitivity are
expected, with the Planck nearing completion.

As the sensitivity of CMB experiments improves, the requirements on
the control and characterization of systematic effects also increase.
It is of critical importance to propagate properly the uncertainties
caused by such effects through to the CMB power spectrum and
cosmological parameters, in order not to underestimate the final
uncertainties, and thereby draw incorrect cosmological conclusions.

A prime example of such systematic effects is non-cosmological
foregrounds in the form of galactic and extra-galactic emission. With
an amplitude rivaling that of the temperature signal over a
significant fraction of the sky and completely dominating the
polarization signal over most of the sky, the diffuse signal from our
own galaxy must be separated accurately from the CMB signal in order
not to bias the cosmological conclusions. Further, the uncertainties
in the separation process must be propagated through to the errors on
the CMB power spectrum and cosmological parameters.

These problems have been discussed extensively in the literature, and
many different approaches to both power spectrum analysis and
component separation have been proposed.  Two popular classes of power
spectrum estimation methods are the pseudo-$C_{\ell}$ estimators
\citep[e.g.,][]{wright:1994, hivon:2002, szapudi:2001} and
maximum-likelihood methods \citep[e.g.,][]{gorski:1994, gorski:1997,
  bond:1998}. For a review and comparison of these methods, see
\citet{efstathiou:2004}. Examples of component separation methods are
the Maximum Entropy Method
\citep{barreiro:2004,bennett:2003b,hobson:1998,stolyarov:2002,stolyarov:2005},
the Internal Linear Combination method
\citep{bennett:2003b,tegmark:2003,eriksen:2004a}, Wiener filtering
\citep{bouchet:1999,tegmark:1996}, the Independent Component Analysis
method \citep{maino:2002,maino:2003,donzelli:2006,stivoli:2006}, and
direct likelihood estimation \citep{brandt:1994,gorski:1996,
  banday:1996, eriksen:2006}.

The final step in a modern cosmological analysis pipeline is
typically to estimate a small set of parameters for some cosmological
model, which in practice is done by mapping out the parameter
posteriors (or likelihoods) using an MCMC code (e.g., CosmoMC; Lewis
and Bridle 2002). To do so, one must establish an expression for the
likelihood $\mathcal{L}(\Cell) = P(\Bd|\Cell)$, where $\Cell$ is a
theoretical CMB power spectrum and $\Bd$ are the observed data. It is
therefore essential that the methods used in the base analysis
pipeline (e.g., map making, component separation, power spectrum
estimation) allow one to estimate this function both accurately and
efficiently.

A particularly appealing framework for this task is the CMB Gibbs
sampler, pioneered by \citet{jewell:2004} and \citet{wandelt:2004}.
While a brute-force CMB likelihood evaluation code must invert a dense
signal-plus-noise covariance matrix, $\BC = \BS + \BN$, at a
computational cost of $\mathcal{O}(\npix^{3})$, $\npix$ being the
number of pixels in the data set, the Gibbs sampler only requires the
signal and noise covariance matrices separately. Consequently, the
algorithmic scaling is dramatically reduced, typically to either
$\mathcal{O}(\npix^{3/2})$ or $\mathcal{O}(\npix^{2})$ for data with
white or correlated noise, respectively. 

In addition to being a highly efficient CMB likelihood evaluator in
its own right, as demonstrated by several previous analyses of real
data \citep{odwyer:2004, eriksen:2007a, eriksen:2007b}, the Gibbs
sampler also offers unique capabilities for propagating systematic
uncertainties end-to-end. Any effect for which there is a well-defined
sampling algorithm, either jointly with or conditionally on other
quantities, can be propagated seamlessly through to the final
posteriors. One example of this is beam uncertainties. Given some
stochastic description of the beam, for instance a mean harmonic space
profile and an associated covariance matrix, one could sample at each
step in the Markov chain one particular realization from this model
and use the resulting beam for the next CMB sampling steps, allowing
for a short burn-in period. The CMB uncertainties will then increase
appropriately. Similar approaches could be taken for uncertainties in
gain calibration and noise estimation.

However, rather than simply propagating a particular error term
through the system, one often wants to estimate the characteristics of
the effect directly from the data. In that case, a parametric model
$P(\theta|\Bd)$, $\theta$ being a set of parameters describing the
effect, must be postulated. Then, if it is both statistically and
computationally feasible to sample from this distribution, the effect
may be included in the joint analysis, and all joint posteriors will
respond appropriately.

In this paper, we describe how non-cosmological frequency-dependent
foreground signals may be included in a Gibbs sampler. In this
framework the CMB signal is assumed Gaussian and isotropic, while the
foregrounds are modeled either in terms of fixed spatial templates
(e.g., monopoles, dipoles, low/high-frequency observations) or in
terms of a free amplitude and spectral response function at each
pixel. Our current code assumes identical angular resolution for all
frequency bands, but we outline in \S\ref{sec:conclusions} how the
algorithm may be generalized to handle multi-resolution experiments.

Already with the present algorithm, we are able to perform a complete
Bayesian joint CMB and foreground analysis of current CMB experiments
on large angular scales. For example, in the present paper we
demonstrate the algorithm on a realistic simulation corresponding to
the 3-yr WMAP data. At an angular resolution of $3^{\circ}$ FWHM, we
are able to produce the exact likelihood up to $\ell \sim 50$-60, into
the regime where a cruder likelihood description is likely to be
acceptable \citep{eriksen:2007a}. Further, in a companion paper
\citep{eriksen:2007c} we analyze the real 3-yr WMAP data with the same
tool, providing for the first time a complete set of physically
motivated foreground posterior distributions of the observed microwave
sky, together with their impact on cosmological parameters.

\section{Review of basic algorithms}
\label{sec:basic_algorithms}

The algorithm developed in this paper is a essentially a hybrid of two
previous algorithms, namely the CMB Gibbs sampler developed by
\citet{jewell:2004}, \citet{wandelt:2004} and \citet{eriksen:2004b},
and the foreground MCMC sampler developed by \citet{eriksen:2006}.  In
this section, we review these algorithms,
emphasizing an intuitive and pedagogical introduction to the
underlying ideas.  In the next section we present the extensions
required to make the hybrid code functional.

Note that while we discuss temperature measurements only in this
paper, the methodology for analyzing polarization measurements is
completely analogous.  For example, see \citet{larson:2007} and
\citet{eriksen:2007b} for details on polarized power spectrum analysis
through Gibbs sampling.

\subsection{The CMB Gibbs sampler}
\label{sec:cmb_sampling}

We first review the Gibbs sampler for CMB temperature measurements.

\subsubsection{The CMB posterior}

We choose our first data model to read
\begin{equation}
\Bd = \Bs + \Bn,
\end{equation}
where $\Bd$ are the observed data, $\Bs$ is the CMB sky signal, and
$\Bn$ is instrumental noise. Complications such as multi-frequency
observations and beam convolution will be introduced at a later stage.

We assume both the CMB signal and noise to be Gaussian random fields
with vanishing mean and covariance matrices $\BS$ and $\BN$,
respectively. In harmonic space, where $\Bs = \sum_{\ell, m} a_{\ell
  m} Y_{\ell m}$, the CMB covariance matrix is given by
$\textrm{C}_{\ell m, \ell' m'} = \langle a_{\ell m}^* a_{\ell'
  m'}\rangle = C_{\ell} \delta_{\ell \ell'} \delta_{m m'}$, $\Cell$
being the angular power spectrum. The noise matrix $\BN$ is left
unspecified for now, but we note that for white noise it is diagonal
in pixel space, $N_{ij} = \sigma_i^2 \delta_{ij}$, for pixels $i$ and
$j$ and noise variance $\sigma^2_i$.

Our goal is to estimate both the sky signal $\Bs$ and the power
spectrum $C_{\ell}$, which in a Bayesian analysis means to compute the
posterior distribution $P(\Bs, C_{\ell}|\Bd)$. By Bayes' theorem, this
distribution may be written as
\begin{align}
P(\Bs, C_{\ell}|\Bd) &\propto P(\Bd|\Bs, C_{\ell}) P(\Bs, C_{\ell}) \\
&\propto P(\Bd|\Bs, C_{\ell}) P(\Bs|C_{\ell}) P(C_{\ell}),
\end{align}
where $P(C_{\ell})$ is a prior on $C_{\ell}$, which we take to be
uniform in the following. Our final power spectrum distribution may
thus be interpreted as the likelihood, and integrated directly into
existing cosmological parameter MCMC codes. Since we have assumed
Gaussianity, the joint posterior distribution may thus be written as
\begin{equation}
P(\Bs, C_{\ell}|\Bd) \propto 
e^{-\frac{1}{2} (\Bd-\Bs)^t \BN^{-1} (\Bd-\Bs)}
\prod_{\ell}\frac{e^{-\frac{2\ell+1}{2}
    \frac{\sigma_{\ell}}{C_{\ell}}}}{C_{\ell}^{\frac{2\ell+1}{2}}}
P(C_{\ell}),
\label{eq:cmb_posterior}
\end{equation}
where $\sigma_{\ell} \equiv \frac{1}{2\ell+1} \sum_{m=-\ell}^{\ell}
|a_{\ell m}|^2$ is the angular power spectrum of the full-sky CMB signal.

\subsubsection{Gibbs sampling}

In principle, we could map out this distribution over a grid in $\Bs$
and $C_{\ell}$, and the task would be done. Unfortunately, since the
number of grid points required for such an analysis scales
exponentially with the number of free parameters, this approach is
not feasible.

A potentially much more efficient approach is to
map out the distribution by sampling. However, direct sampling from
the joint distribution in Equation \ref{eq:cmb_posterior} is difficult
even from an algorithmic point of view alone; we are not aware of any
textbook approach for this. And even if there were, it would most
likely involve inverses of the joint $\BS + \BN$ covariance matrix,
with a prohibitive $\mathcal{O}(\npix^3)$ scaling, in order to
transform to the eigenspace of the system.

This is the situation in which \citet{jewell:2004} and
\citet{wandelt:2004} proposed a particular Gibbs sampling scheme. For
a general introduction to the algorithm, see, e.g.,
\citet{gelfand:1990}. 
In short, the theory of Gibbs sampling tells
us that if we want to sample from the joint density $P(\Bs,
C_{\ell}|\Bd)$, we can alternately sample from the respective
conditional densities as follows,
\begin{align}
\Bs^{i+1} &\leftarrow P(\Bs | C_{\ell}^i, \Bd) \\
C_{\ell}^{i+1} &\leftarrow P(C_{\ell} | \Bs^{i+1}, \Bd).
\end{align}
Here $\leftarrow$ indicates sampling from the distribution on the
right-hand side. After some burn-in period, during which the samples
must be discarded, the joint samples $(\Bs^i, C_{\ell}^i)$ will be
drawn from the desired density. Thus, the problem is reduced to that
of sampling from the two \emph{conditional} densities $P(\Bs |
C_{\ell}, \Bd)$ and $P(C_{\ell} | \Bs, \Bd)$.

\subsubsection{Sampling algorithms for conditional distributions}
\label{sec:cond_samp_algorithms}

We now describe the sampling algorithms for each of these two
conditional distributions, starting with $P(C_{\ell} | \Bs, \Bd)$.
First, note that $P(C_{\ell} | \Bs, \Bd) = P(C_{\ell} | \Bs)$; if we
already know the CMB sky signal, the data themselves tell us nothing
new about the CMB power spectrum. Next, since the sky is assumed
Gaussian and isotropic, the distribution reads
\begin{equation}
P(C_{\ell} | \Bs) \propto \frac{e^{-\frac{1}{2}
    \Bs_{\ell}^{t}\BS_{\ell}^{-1}\Bs_{\ell}}}{\sqrt{|\BS_{\ell}|}} =    \frac{e^{-\frac{2\ell+1}{2} \frac{\sigma_{\ell}}{C_{\ell}}}}{C_{\ell}^{\frac{2\ell+1}{2}}},
\end{equation}
which, when interpreted as a function of $C_{\ell}$, is known as the
inverse Gamma distribution. Fortunately, there exists a simple
textbook sampling algorithm for this distribution
\citep[e.g.,][]{eriksen:2004b}, and we refer the interested reader to
the previous papers for details.

The sky signal algorithm is even simpler from a statistical point of
view, although more involved to implement.  Defining the so-called
mean-field map (or Wiener filtered data) to be $\hat{\Bs} = (\BS^{-1}
+ \BN^{-1})^{-1} \BN^{-1} \Bd$, the conditional sky signal distribution may
be written as
\begin{align}
P(\Bs | C_{\ell}, \Bd) &\propto P(\Bd|\Bs, C_{\ell}) P(\Bs|C_{\ell}) \\
&\propto e^{-\frac{1}{2} (\Bd-\Bs)^t \BN^{-1} (\Bd-\Bs)} 
\,\,e^{-\frac{1}{2}\Bs^t\BS^{-1}\Bs} \\
&\propto e^{-\frac{1}{2} (\Bs-\hat{\Bs})^t (\BS^{-1} + \BN^{-1}) (\Bs-\hat{\Bs})}.
\end{align}
Thus, $P(\Bs | C_{\ell}, \Bd)$ is a Gaussian distribution with mean
equals to $\hat{\Bs}$ and a covariance matrix equals to $(\BS^{-1} +
\BN^{-1})^{-1}$.

Sampling from this Gaussian distribution is straightforward, but
computationally somewhat cumbersome. First, draw two random white
noise maps $\omega_0$ and $\omega_1$ with zero mean and unit
variance. Then solve the equation
\begin{equation}
\left[\BS^{-1} + \BN^{-1}\right] \Bs = \BN^{-1}\Bd + \BS^{-\frac{1}{2}} \omega_0 +
\BN^{-\frac{1}{2}} \omega_1.
\label{eq:lin_sys}
\end{equation}
for $\Bs$. Since the white noise maps have zero mean, one immediately
sees that $\langle \Bs \rangle = \hat{\Bs}$, while a few more
calculations show that $\langle \Bs \Bs^{t} \rangle = (\BS^{-1} +
\BN^{-1})^{-1}$. 

The problematic part about this sampling step is the solution of the
linear system in Equation \ref{eq:lin_sys}. Since this a $\sim10^6
\times 10^6$ system for current CMB data sets, it cannot be solved by
brute force. Instead, one must use a method called Conjugate Gradients
(CG), which only requires multiplication of the coefficient matrix on
the left-hand side, not inversion. For details on these computations,
together with some ideas on preconditioning, see
\citet{eriksen:2004b}.

\subsubsection{Generalization to multi-frequency data}

For notational transparency, the discussion in the previous sections was
limited to analysis of a single sky map, and did not
include the effect of an instrumental beam. We now review the full
equations for the general case. See \citet{eriksen:2004b} for full
details.

Let $\Bd_{\nu}$ denote an observed sky map at frequency $\nu$,
$\BN_{\nu}$ its noise covariance matrix, and $\BA_{\nu}$ convolution
with the appropriate instrumental beam response. Equation
\ref{eq:lin_sys} then generalizes to
\begin{equation}
\begin{split}
\left[\BS^{-1} + \sum_{\nu} \BA_{\nu}^t \BN_{\nu}^{-1}\BA_{\nu}\right]
\Bs = & \\ \sum_{\nu} \BA_{\nu}^t \BN_{\nu}^{-1}\Bd_{\nu}
+ \BS^{-\frac{1}{2}} &\omega_0 + \sum_{\nu} \BA_{\nu}^t
\BN_{\nu}^{-\frac{1}{2}} \omega_{\nu}.
\end{split}
\label{eq:lin_sys2}
\end{equation}
Note that we now draw one white noise map for each frequency band,
$\omega_{\nu}$. The sampling procedure for $P(C_{\ell}|\Bs)$ is
unchanged.

\subsubsection{Computational considerations}

Finally, we make two comments regarding numerical stability and
computational expense. First, note that the elements of $\Bs$ have a
variance equal to the CMB power spectrum, which goes as $C_{\ell} \sim
\ell^{-2}$.  To avoid round-off errors over the large dynamic range in
the solution, it is numerically advantageous to solve first 
for $\mathbf{x} = \BS^{-\frac{1}{2}}\Bs$ in the CG search, and
then to solve (trivially) for $\Bs$. The system solved by
CG in practice is thus 
\begin{equation}
\begin{split}
\left[\mathbf{1} + \BS^{\frac{1}{2}}\sum_{\nu} \BA_{\nu}^t
\BN_{\nu}^{-1}\BA_{\nu}\BS^{\frac{1}{2}}\right]
\Bx &= \\ \BS^{\frac{1}{2}} \sum_{\nu}
\BA_{\nu}^t \BN_{\nu}^{-1}\Bd_{\nu} + \omega_0 +
&\BS^{\frac{1}{2}} \sum_{\nu} \BA_{\nu}^t \BN_{\nu}^{-\frac{1}{2}} \omega_{\nu}.
\end{split}
\label{eq:lin_sys3}
\end{equation}

Second, solving this equation by CG involves multiplication with
expression in the brackets on the left-hand side, and therefore scales
as the most expensive operation in the coefficient matrix.  For white
noise, $\BN_{ij} = \sigma_i \delta_{ij}$, this is the spherical
harmonic transform required between pixel (for noise covariance matrix
multiplication) and harmonic (for beam convolution and signal
covariance matrix multiplication) space, with a scaling of
$\mathcal{O}(\npix^{3/2})$. For correlated noise, it is the
multiplication with a dense $\npix \times \npix$ inverse noise
covariance matrix, with a scaling of $\mathcal{O}(\npix^{2})$.

\subsection{The foreground sampler}
\label{sec:fg_sampling}

The previous section described how to sample from the exact CMB
posterior $P(\Bs, C_{\ell}|\Bd)$ by Gibbs sampling. In this section,
we very briefly review the algorithm for sampling general sky signals
presented by \citet{eriksen:2006}.

First we define a parametric frequency model for the total sky signal,
$S_{\nu}(\theta)$, $\theta$ representing the set of all free
parameters in the model. A simple example would be
$S(T_{\textrm{cmb}}, A_{\textrm{s}}, \beta_{\textrm{s}}) =
T_{\textrm{cmb}} + A_{\textrm{s}} a(\nu)
\left(\frac{\nu}{\nu_0}\right)^{\beta_{\textrm{s}}}$, where
$T_{\textrm{cmb}}$ is the CMB temperature, $A_{\textrm{s}}$ is the
synchrotron emission amplitude relative to a reference frequency
$\nu_0$, $\beta_{\textrm{s}}$ is the synchrotron spectral index, and
$a(\nu)$ is the conversion factor between antenna and
thermodynamic temperature for differential measurements. Note that no
constraints are imposed on the form of the spectral model in general,
beyond the fact that it should contain at most $N_{\nu}-1$ free
parameters, $N_{\nu}$ being the number of frequency bands of the
experiment. In practice, one should also avoid models that contain
nearly degenerate parameters.

Our goal is now to compute the posterior distribution $P(\theta|\Bd)$
for each pixel. For this to be computationally feasible, we make two
assumptions. First, we assume that the noise is uncorrelated between
pixels, and second, that the instrumental beams are identical between
frequency bands. If so, the data may be analyzed pixel-by-pixel, and
the likelihood for a single pixel simply reads
\begin{equation}
-2\ln\mathcal{L}(\theta) = \chi^2(\theta) = \sum_{\nu}
\left(\frac{d_{\nu}-S_{\nu}(\theta)}{\sigma_{\nu}}\right)^2.
\label{eq:chisq}
\end{equation}
The posterior is as usual given by $P(\theta|\Bd) \propto
\mathcal{L}(\theta) P(\theta)$, where $P(\theta)$ is a prior on
$\theta$. Given this likelihood and prior, it is straightforward to
sample from $P(\theta|\Bd)$, for instance by Metropolis-Hastings MCMC
\citep{eriksen:2006}, or by inversion sampling as described later in
this paper.

\section{Joint CMB and foreground sampling}
\label{sec:hybrid_sampling}

The main goal of this paper is to merge the two algorithms described
in \S\S\ref{sec:cmb_sampling} and \ref{sec:fg_sampling} into one
joint CMB--foreground sampler, allowing us to estimate the joint
posterior $P(\Bs, C_{\ell}, \theta|\Bd)$. In this paper we focus on a
matched beam response experiment, for which $\BA_{\nu} = \BA$, which
is sufficient for low-resolution analysis of high-resolution
experiments such as WMAP and Planck. 

\subsection{Data model and priors}
\label{sec:data_model}

We define the joint data model to be 
\begin{equation}
\begin{split}
\Bd_{\nu} &= \BA \Bs + \sum_{i=1}^{M} a_{\nu, i} \mathbf{t}_{i}
+ \sum_{j=1}^{N} b_{j} f_j(\nu) \mathbf{f}_{j} + \\ &\quad\quad+
\sum_{k=1}^{K} \mathbf{c}_{k} \mathbf{g}_{k}(\nu; \theta_k)+ \Bn_{\nu}.
\end{split}
\label{eq:model}
\end{equation}
The first term on the right-hand side is the CMB sky signal. The
second term is a sum over $M$ spatial templates, $\mathbf{t}_{i}$,
each having a free amplitude $a_{\nu,i}$ at each frequency, for example
monopole and dipole components. The third
term is a sum over $N$ spatial templates, $\mathbf{f}_j$ with a fixed 
frequency scaling $f_j(\nu)$
and a single overall amplitude $b_j$, for example the H$_\alpha$ 
template \citep[e.g.,][]{dickinson:2003} coupled
to a power law spectrum with free-free spectral index of
$\beta_{\textrm{ff}}=-2.15$.  Such spatial templates are a way of
incorporating constraints on the sky from other measurements.  
Their value depends on the validity of assumptions about spectra
over large frequency ranges (e.g., H$_\alpha$ as a proxy for free-free
emission at CMB frequencies).  Nevertheless, CMB experiments with
too few frequencies to constrain foregrounds adequately on their own require
such templates to provide additional constraints.  Both
$\mathbf{t}_{i}$ and $\mathbf{f}_{j}$ are assumed to be convolved to
the appropriate angular resolution of the experiment.   

The fourth term, the most important novel feature of this paper, is a 
sum over $K$ foreground components each given by an overall 
amplitude $c_k(p)$ and a frequency spectrum $g_k(\nu; \theta_k(p))$ at
each pixel $p$.   The spectral parameters $\theta_k(p)$ may or may not be allowed to vary
from pixel to pixel.  By allowing independent frequency spectra
at every single pixel, the model is very general, and capable of
describing virtually any conceivable sky signal.

The fifth and last term, $\Bn_{\nu}$, is instrumental noise.

In the current implementation of our codes, we allow only foreground
spectra parametrized by a single spectral index, $g(\nu; \beta) =
\mathcal{G}(\nu) (\nu/\nu_0)^{\beta}$, where $\mathcal{G}(\nu)$ is an
arbitrary, but fixed, function of frequency and $\nu_0$ is a reference
frequency. A typical example is synchrotron emission, which may be
modelled accurately over a wide frequency range by a
simple power law (in intensity, flux density, or antenna temperature units) 
with a spatially varying spectral index.  The CMB is most naturally
described in terms of thermodynamic units, and we adopt this
convention in our codes. The corresponding synchrotron model is
therefore $g(\nu; \beta) = a(\nu) (\nu/\nu_0)^{\beta}$, where $a(\nu)$
is the antenna-to-thermodynamic conversion factor.


As in any Bayesian analysis, we must adopt a set of priors
for the parameters under consideration.  For this
paper, we choose the prior most widely accepted in the
statistical community, namely Jeffreys' ignorance prior
\citep[e.g.,][]{box:1992}. This prior is given by the square root of
the Fisher information measure, $P_{\theta} \propto
\sqrt{|F|_{\theta\theta}} = \sqrt{|-\partial^2 \ln \mathcal{L} /
  \partial^2\theta|}$. Its effect is essentially to ``normalize'' the
parameter volume relative to the likelihood, and make the likelihood
so-called ``data translated''. We will return to the effect of this
prior in \S\ref{sec:flat_vs_jeffrey}.  We impose an additional
multiplicative prior on spectral parameters, either a top-hat or a Gaussian.

For amplitude-type degrees of freedom, the ignorance prior works out
be the usual flat prior, but for non-linear parameters, e.g., spectral
indices, it is non-uniform. In particular, for the power law spectrum
parametrized by a spectral index $\beta$ described above, it reads
$P(\beta) \propto \sqrt{\sum_{\nu} ((\mathcal{G}(\nu)/\sigma_{\nu})
  (\nu/\nu_0)^{\beta} \ln (\nu/\nu_0))^2}$. The difference between
this and a flat prior will be demonstrated in \S\ref{sec:flat_vs_jeffrey}.

An important special case is the CMB power spectrum, for which we
adopt a uniform prior despite the fact that the corresponding density
is non-Gaussian. The main reason for doing so is that most
cosmological parameter estimation codes expect the CMB likelihood,
rather than the CMB posterior.  

\subsection{Sampling from the joint posterior distribution}

Having defined our data model and priors, the goal is now to estimate
the joint CMB--foreground posterior $P(\Bs, C_{\ell}, a_{\nu,i}, b_j,
\mathbf{c}_k, \theta_k|\Bd)$. This is achieved through the following
straightforward generalization of the previous Gibbs sampling scheme,
\begin{align}
\{\Bs, a_{\nu,i}, b_j, \mathbf{c}_k\}^{i+1} &\leftarrow P(\Bs, a_{\nu,i},
b_j, \mathbf{c}_k | C_{\ell}^i, \theta_k^i, \Bd) \label{eq:amplitude_sampler}\\ 
\theta_k^{i+1} &\leftarrow P(\theta_k |\Bs^{i+1}, a_{\nu,i}^{i+1}, b_j^{i+1}, \mathbf{c}_k^{i+1}, \Bd) \label{eq:theta_sampling}\\
C_{\ell}^{i+1} &\leftarrow P(C_{\ell} | \Bs^{i+1})\label{eq:spectrum_sampler}.
\end{align}
Explicitly, all amplitude-type degrees of freedom are sampled jointly
with the CMB sky signal using a generalization of Equation
\ref{eq:lin_sys}, while all non-linear spectral parameters are sampled
conditionally by inversion sampling, as described in \S\ref{sec:index_sampling}. The conditional CMB power spectrum sampling
algorithm is unchanged, since it only depends on the CMB sky signal.

\subsubsection{Amplitude sampling}
\label{sec:amplitude_sampling}

We first describe the algorithm for sampling from the conditional
amplitude density, $P(\Bs, a_{\nu,i}, b_j, \mathbf{c}_k | C_{\ell},
\theta_k, \Bd)$

\paragraph{Conditional sampling of amplitudes}
\label{sec:conditional_sampling}
In principle, we could take further advantage of the Gibbs sampling
approach, and sample each of $\Bs$, $a_{\nu, i}$, $b_j$ and
$\mathbf{c}_k$ conditionally, given all other parameters, including
the amplitudes not currently being sampled. This method was briefly
described by \citet{eriksen:2004b} for monopole and dipole sampling,
and later used for actual analysis by both \citet{odwyer:2004} and
\citet{eriksen:2007b}. Briefly stated, this approach simply amounts to
subtracting each of the signals that is conditioned upon from the
data, and using the residual map to sample the remaining parameters in
place of the full data set. Its main advantage is highly modularized,
simple and transparent computer code.

However, for general applications this is a prohibitively inefficient
sampling algorithm due to poor mixing properties and long Markov chain
correlation lengths. The problem is due to strong correlations between
the various amplitudes. Consider for instance a model including a CMB
sky signal, monopole and dipole components, and a foreground
template. Note that the latter has both a non-zero monopole and dipole
and also smaller-scale structure.

The conditional sampling algorithm would then go as follows: First
subtract the current monopole, dipole and foreground components from
the data, and sample the CMB sky based on the residual map. The
uncertainties in this conditional distribution are both cosmic
variance and instrumental noise. Second, subtract the recently sampled
CMB signal and foreground template from the data, and sample the
monopole and dipoles of the residual. The only source of uncertainty
in this conditional distribution is instrumental noise alone, and the
next sample therefore equals the previous state plus a noise
fluctuation.  For high signal-to-noise data the instrumental noise
uncertainty in a single all-sky number such as the monopole and dipole
amplitude is very small indeed, and the new sample is therefore
essentially identical to the previous. Finally, subtract the CMB
signal and the monopole and dipole from the data, and sample the
foreground template amplitude of the new residual. Again, with high
signal-to-noise data the new amplitude is virtually identical to the
previous.

The failure of this approach stems from the fact that the main
uncertainty in the monopole, dipole and template amplitudes is not
instrumental noise, but rather CMB cosmic variance coupled from
template structures.  This component is not explicitly acknowledged in
the conditional template sampling algorithms when conditioning on the
CMB signal, but only implicitly through the Gibbs sampling chain.  The
net result is an extremely long Markov chain correlation length.

The reasons this conditional approach worked well in the analyses of
\citet{eriksen:2004b}, \citet{odwyer:2004} and \citet{eriksen:2007b}
cases were different, and somewhat fortuitous: Only the monopole and
dipole components were included in the 1-yr WMAP temperature analysis,
which couple only weakly to the low-$\ell$ CMB modes with a relatively
small sky-cut. No foreground template sampling step as such was
included, which would couple strongly to both the monopole, dipole and
CMB signals. For the polarization analysis of \citet{eriksen:2007b},
in which foreground templates were indeed included, a different effect
came into play, namely the very low signal-to-noise ratio of the 3-yr
WMAP polarization data. At this signal-to-noise, even conditional
sampling works well.

\paragraph{Joint sampling of amplitudes}
The solution to this problem is to sample all amplitude-type degrees
of freedom jointly from $P(\Bs, a_{\nu,i}, b_j, \mathbf{c}_k |
C_{\ell}, \theta_k, \Bd)$. This is a four-component Gaussian
distribution with mean $\hat{\mathbf{x}}$ and covariance matrix
$\mathcal{A}$. The required sampling algorithm is therefore fully
analogous to that described in \S\ref{sec:cond_samp_algorithms} 
for the CMB sky signal. The remaining
task is to generalize the expressions for $\hat{\mathbf{x}}$ and
$\mathcal{A}$.

To keep the notation tractable, we first define a symbolic
four-element block vector of all amplitude coefficients, $\mathbf{x} =
(\Bs, a_{\nu,i}, b_j, \mathbf{c}_k)^t$. The first block of
$\mathbf{x}$ contains the harmonic coefficients of $\Bs$, the second
block contains $a_{\nu,i}$ for all frequencies and templates, the
third contains $b_j$ for all templates with a fixed spectrum, and the
fourth contains the pixel amplitudes $\mathbf{c}_k$ for all
pixel-by-pixel foreground components. In total, $\mathbf{x}$ is an
$((\ell_{\textrm{max}}+1)^2 + M \,N_{\textrm{band}} + N + K\,
\npix)$-element vector. We also define a corresponding response vector
$\mathbf{u}_{\nu} = (\mathbf{1}, \mathbf{t}_i, f_{j}(\nu)\mathbf{f}_j,
\mathbf{g}_{k}(\nu; \theta_k))^t$, such that the data model in
Equation \ref{eq:model} may be abbreviated to $\Bd_{\nu} = \mathbf{x}
\cdot \mathbf{u}_{\nu} + \Bn_{\nu}$.

With this notation, the joint amplitude distribution reads
\begin{equation*}
\begin{split}
P(\Bs, a_{\nu,i}, b_j, \mathbf{c}_k | C_{\ell}, \theta_k, \Bd)
\propto P(\Bd|\Bs, C_{\ell},&a_{\nu,i}, b_j, \mathbf{c}_k)
P(\Bs|C_{\ell}) \\
\propto e^{-\frac{1}{2} \sum_{\nu}(\Bd_{\nu}-\Bx\cdot\Bu_{\nu})^t \BN_{\nu}^{-1}
  (\Bd_{\nu}-\Bx\cdot\Bu_{\nu})} 
\,\, &e^{-\frac{1}{2}\Bs^t\BS^{-1}\Bs} \\
\propto e^{-\frac{1}{2} (\Bx-\hat{\Bx})^t \mathcal{A}^{-1}
  (\Bx-\hat{\Bx})}.\quad\quad\quad\quad\,\,&
\end{split}
\end{equation*}
Here we have implicitly defined the symbolic $4\times4$ inverse
covariance block matrix
\begin{equation}
\mathcal{A}^{-1} = \left[ \begin{array}{cccc}
\BS^{-1} +  \BA^{t} \BN^{-1} \BA  &  \BA^{t} \BN^{-1} \BT & \BA^{t} \BN^{-1} \BF & \BA^{t}
\BN^{-1} \BG \\
\BT^{t} \BN^{-1} \BA & \BT^{t} \BN^{-1} \BT & \BT^{t} \BN^{-1} \BF & \BT^{t} \BN^{-1} \BG \\
\BF^{t} \BN^{-1} \BA & \BF^{t} \BN^{-1} \BT & \BF^{t} \BN^{-1} \BF & \BF^{t} \BN^{-1} \BG \\
\BG^t \BN^{-1} \BA & \BG^t \BN^{-1} \BT & \BG^t \BN^{-1} \BF & \BG^t \BN^{-1} \BG
\end{array} \right]
\label{eq:coupling_matrix}
\end{equation}
(see Appendix \ref{app:coupling_matrix} for explicit definitions of
each element in this matrix) and a corresponding four-element symbolic
block vector for the Wiener-filter mean,
\begin{equation}
\hat{{\bf x}} = \mathcal{A} \left[ \begin{array}{c}
\sum_{\nu} \BA^{t} \BN_{\nu}^{-1} \Bd \\
\mathbf{t}^{t}_{\nu,j} \BN_{\nu}^{-1} \Bd \\
\sum_{\nu} f_j(\nu) \mathbf{f}_j^{t} \BN^{-1} \Bd \\
\sum_{\nu} \mathbf{g}_k(\nu; \theta_k) \BN_{\nu}^{-1} \Bd
\end{array} \right].
\label{eq:joint_mean}
\end{equation}

The sampling algorithm for this joint distribution is now fully
analogous to the one described in \S\ref{sec:cond_samp_algorithms} 
by Equation \ref{eq:lin_sys2}: 1) Draw
$N_{\textrm{band}}+1$ white noise maps with zero mean and unit
variance; 2) form the Wiener filter mean plus random fluctuation
right-hand side vector,
\begin{equation}
{\bf b} = \left[ \begin{array}{c}
\sum_{\nu} \BA^{t} \BN_{\nu}^{-1} \Bd + \BC^{\frac{1}{2}} \omega_0 +
\sum_{\nu} \BA_{\nu}^t \BN_{\nu}^{-\frac{1}{2}} \omega_{\nu}\\
\mathbf{t}^{t}_{\nu,j} \BN_{\nu}^{-1} \Bd + \mathbf{t}^{t}_{\nu,j} \BN_{\nu}^{-\frac{1}{2}} \omega_{\nu}\\
\sum_{\nu} f_j(\nu) \mathbf{f}_j^{t} \BN^{-1} \Bd + \sum_{\nu} f_j(\nu) \mathbf{f}_j^{t} \BN^{-\frac{1}{2}} \omega_{\nu} \\
\sum_{\nu} \mathbf{g}_k(\nu; \theta_k) \BN_{\nu}^{-1} \Bd + \sum_{\nu} \mathbf{g}_k(\nu; \theta_k) \BN_{\nu}^{-\frac{1}{2}} \omega_{\nu}
\end{array} \right];
\end{equation}
and 3) solve the set of linear equations,
\begin{equation}
\mathcal{A}^{-1}{\bf x} = {\bf b}.
\label{eq:gen_lin_sys}
\end{equation}
The solution vector ${\bf x}$ then has the required mean $\hat{{\bf
x}}$ and covariance matrix $\mathcal{A}$. Again, for numerical
stability it is useful to multiply both sides of Equation
\ref{eq:gen_lin_sys} by the block-diagonal matrix $\mathbf{P} =
\textrm{diag}(\BC^{-\frac{1}{2}}, \mathbf{1}, \mathbf{1},\mathbf{1})$,
and solve for $\mathbf{P}\mathbf{x}$ by CG.

\begin{figure}[t]
\mbox{\epsfig{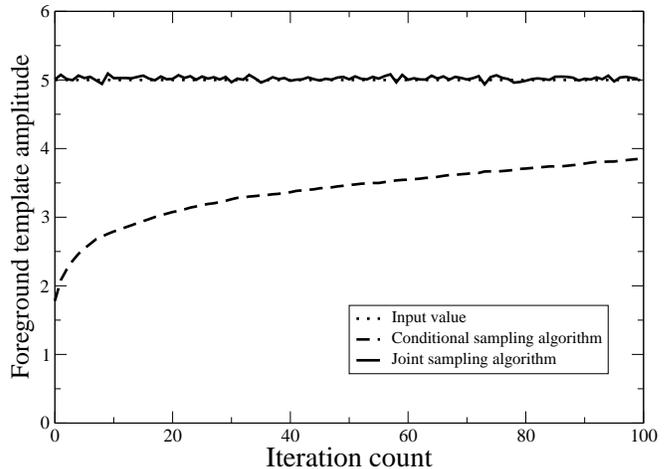}}
\caption{Comparison of trace plots generated by the joint (solid
  curve) and the conditional (dashed curve) template amplitude
  sampling algorithms for a simulated data set consisting of a CMB sky
  signal, a monopole component and a synchrotron template component.
  The true input template amplitude is shown as a horizontal dotted
  line.}
\label{fig:joint_vs_conditional}
\end{figure}

To demonstrate the difference in mixing efficiency between conditional
and joint amplitude sampling, Figure\,\ref{fig:joint_vs_conditional} shows 
two trace plots for a high
signal-to-noise simulation that included a CMB, a monopole, and a
foreground template component. While the joint sampler instantaneously
moves into the right regime, and subsequently efficiently explores the
correct distribution, the conditional sampler converges only very
slowly toward the correct value. The associated long Markov chain
correlation length makes this approach unfeasible for general
problems.

\paragraph{Preconditioning}
\label{sec:preconditioning}
The performance of the CG algorithm (see \citet[][]{shewchuk:1994} for an
outstanding introduction to this method) depends sensitively on the
condition number of the coefficient matrix $\mathcal{A}$, i.e., the
ratio of the largest to the smallest eigenvalue. In fact, the
algorithm is not guaranteed to converge at all for poorly conditioned
matrices, due to increasing round-off errors in cases that require
many iterations.

The condition number of the regularized $\mathcal{A}$ matrix is
essentially the largest signal-to-noise ratio of any component in the
system, which in practice means that of the CMB quadrupole or the
template amplitudes. For current and future CMB experiments, such as
WMAP and Planck, the integrated signal-to-noise of these large-scale
modes is very large. It is therefore absolutely essential to construct
an efficient preconditioner, $\mathbf{M} \approx \mathcal{A}$, to
decouple these modes brute-force, $\mathbf{M}^{-1} \mathcal{A} \Bx =
\mathbf{M}^{-1}\mathbf{b}$, simply in order to achieve basic
convergence.

For the $4\times4$ coupled system described above, we adopt a
three-stage preconditioner. First, for the low-$\ell$ CMB components
we explicitly compute all elements of $\mathcal{A}$ up to some
$\ell_{\textrm{precond}} \sim 20$--70 \citep{eriksen:2004b}. This
low-$\ell$ block is then coupled to the template amplitudes in a
symbolic $3\times3$ preconditioner,
\begin{equation}
\mathbf{M}_0 = \left[ \begin{array}{ccc}
\BS^{-1} +  \BA^{t} \BN^{-1} \BA  &  \BA^{t} \BN^{-1} \BT & \BA^{t} \BN^{-1} \BF \\
\BT^{t} \BN^{-1} \BA & \BT^{t} \BN^{-1} \BT & \BT^{t} \BN^{-1} \BF  \\
\BF^{t} \BN^{-1} \BA & \BF^{t} \BN^{-1} \BT & \BF^{t} \BN^{-1} \BF  
\end{array} \right].
\label{eq:preconditioner}
\end{equation}
The elements in this matrix are computed by transforming each object
individually into spherical harmonic space, including modes only up to
$\ell_{\textrm{precond}}$, and then performing the sums explicitly.
(Note that the seemingly intuitive proposition of computing the
template elements in pixel space, as opposed to in harmonic space, is
flawed; unless all elements are properly bandwidth limited, a
non-positive definite preconditioning matrix will result.) For
examples of such computations, see \citet{eriksen:2004b}.

The second part of our preconditioner regularizes the high-$\ell$ CMB
components, and consists of the diagonal elements $\mathcal{A}_{\ell
  m, \ell' m'} \delta_{\ell \ell'} \delta_{m m'}$ from
$\ell_{\textrm{precond}}+1$ to $\ell_{\textrm{max}}$
\citep{eriksen:2004b}.  The third part of our preconditioner covers 
the single pixel-pixel foreground amplitudes, which
have low signal-to-noise ratio,  and are preconditioned with the 
corresponding diagonal elements of
$\mathcal{A}$ only, $\BG^t \BN^{-1} \BG$.



For a typical low-resolution WMAP3 application (five frequency
channels degraded to $N_{\textrm{side}}=64$ and $3^{\circ}$ FWHM
resolution and regularized with $2\mu\textrm{K}$ RMS white noise), we
find that including only the diagonal elements in the above matrix can
bring the fractional CG residual down to $\sim 10^{-4}$, while the
recommended convergence criterion for single-precision data is
$10^{-6}$. Thus, including the CMB-template cross-terms in the
low-$\ell$ CMB preconditioner in Equation \ref{eq:preconditioner} is
not just a question of performance for the signal-to-noise levels of
WMAP; it is required in order to converge at all. The total number of
CG iterations is typically $\lesssim 200$ for the same application
with the previously described three-level preconditioner. For some
further promising ideas on preconditioning for similar systems, see
\citet{smith:2007}.

\paragraph{Imposing linear constraints}
\label{sec:constraints}
A useful addition to the above formalism is the possibility of
imposing linear constraints on one or more of the parameters. For
instance, if it is possible to calibrate the absolute offset of one
frequency band by external information, for instance using knowledge
about the instrument itself, it would be highly beneficial to fix the
corresponding monopole value accordingly. Another constraint may be to
exclude template amplitude combinations with a given frequency
spectrum, in order to disentangle arbitrary offsets at each frequency
from the absolute zero-level of a given foreground component.

In the present code, we have implemented an option for imposing linear
constraints on the template amplitudes $\mathbf{a} = \{a_{\nu,i}\}$ on the
form
\begin{equation}
  \sum_{\nu, i} q^k_{\nu,i} a_{\nu,i} = \mathbf{q}^k\cdot\mathbf{a} =
  0,
\label{eq:constraints}
\end{equation}
where $\mathbf{q}^k = \{q^k_{\nu,i}\}$, $k = 1,\ldots,N_{\textrm{c}}$,
are constant orthogonality vectors, and $N_{\textrm{c}}$ is the number
of simultaneous linear constraints. For example, if we want to obtain
a solution with a fixed monopole amplitude at frequency $\nu_0$, we
would set $q_{\nu,i} = \delta_{\nu,\nu_0} \delta_{i,0}$.

The total dimension of the template amplitude vector space is $D =
MN_{\textrm{band}}$, $M$ being the number of free templates at each
band. Within this space, the constraint vectors $\mathbf{q}$ span an
$N_{\textrm{c}}$-dimensional sub-space $\mathcal{V}$ to which the CG
solution must be orthogonal; $\mathbf{a}$ must lie in the complement
of $\mathcal{V}$, denoted $\mathcal{V}^{\textrm{c}}$.

To achieve this, we construct a projection operator $P: R^{D}
\rightarrow \mathcal{V}^{\textrm{c}}$ by standard Gram-Schmidt
orthogonalization, which is a $D\times
(D-N_{\textrm{c}})$-dimensional matrix $\mathbf{P}$.  To impose the
constraints defined by Equation \ref{eq:constraints} on the the final
CG solution, Equation \ref{eq:gen_lin_sys} is rewritten as
\begin{equation}
  \mathbf{P}^{\textrm{t}}\mathcal{A}^{-1}\mathbf{P} \mathbf{P}^\textrm{t}{\bf x} = \mathbf{P}^{t}{\bf b},
\end{equation}
which is solved as before. Corresponding elements in the
preconditioner are similarly modified in order to maintain
computational efficiency.

\subsubsection{Spectral parameter sampling}
\label{sec:index_sampling}

With the amplitude sampling equations for $P(\Bs,
a_{\nu,i}, b_j, \mathbf{c}_k | C_{\ell}, \theta_k, \Bd)$ in hand, the only
missing piece in the Gibbs sampling scheme defined by Equations
\ref{eq:amplitude_sampler}--\ref{eq:spectrum_sampler}, is a spectral
parameter sampler for $P(\theta_k |\Bs, a_{\nu,i}, b_j, \mathbf{c}_k,
\Bd)$. In the FGFit code presented by \citet{eriksen:2006}, this task
was done by Metropolis-Hastings MCMC, a very general
technique that can sample from almost any multi-variate distribution.
However, it has two disadvantages.  First, hundreds of MCMC steps 
may be required to
generate two uncorrelated samples, making the process quite expensive.
Second, and even worse for our application, the chains may need to
``burn in'' at each main Gibbs iteration, because the amplitude
parameters have changed since the last iteration. Proper monitoring of
these issues is difficult for problems with tens of thousands of
pixels with very different signal-to-noise ratios.

Therefore, we have replaced the MCMC sampler with a direct
sampler, specifically a standard inversion sampler, in the
present version of our codes. While this algorithm is only applicable
for univariate problems, it is also quite possibly the best such
sampler, as it draws from the exact distribution, and no computation
of acceptance probabilities is needed. The algorithm is the
following: First, compute the conditional probability density
$P(x|\theta)$, where $x$ is the currently sampled parameter and
$\theta$ denotes the set of all other parameters in the model. In our
application, $P(x|\theta)$ is the normalized product of the likelihood
$\mathcal{L}$ in Equation \ref{eq:chisq} and any prior we wish to
impose. Then compute the corresponding cumulative probability
distribution, $F(x|\theta) = \int_{-\infty}^{x} P(y|\theta) dy$. Next,
draw a random number $u$ from the uniform distribution $U[0,1]$. The
desired sample from $P(x|\theta)$ is given by $F(x|\theta) = u$.

For multivariate problems we use a Gibbs sampling scheme to draw from
the joint distribution, and sample each parameter conditionally. For
example, if we want to allow free amplitudes ($c_{\textrm{s}}$ and
$c_{\textrm{d}}$) and spectral indices ($\beta_{\textrm{s}}$ and
$\beta_{\textrm{d}}$) for both synchrotron and thermal dust emission,
the full sampling scheme reads
\begin{align}
\{\Bs, \mathbf{c}_{\textrm{s}}, \mathbf{c}_{\textrm{s}}\}^{i+1}
&\leftarrow P(\Bs, \mathbf{c}_{\textrm{s}}, \mathbf{c}_{\textrm{s}} |
C_{\ell}^i, \beta_{\textrm{s}}^i, \beta_{\textrm{s}}^i, \Bd)\\ 
C_{\ell}^{i+1} &\leftarrow P(C_{\ell} | \Bs^{i+1})\\
\beta_{\textrm{s}}^{i+1} &\leftarrow P(\beta_{\textrm{s}} |\Bs^{i+1},
,\mathbf{c}_{\textrm{s}}^{i+1}, \mathbf{c}_{\textrm{s}}^{i+1},
\beta_{\textrm{d}}^{i}, \Bd)\\
\beta_{\textrm{d}}^{i+1} &\leftarrow P(\beta_{\textrm{d}} |\Bs^{i+1},
,\mathbf{c}_{\textrm{s}}^{i+1}, \mathbf{c}_{\textrm{s}}^{i+1}, \beta_{\textrm{s}}^{i}, \Bd)
\end{align}
Note that it can be beneficial to iterate the latter two equations
more than once in each main Gibbs loop, in order to reduce the
correlations between consequtive samples cheaply. Typically, with two
moderately correlated spectral indices we run $\sim3$ spectral index
iterations for each main Gibbs iteration.

While this approach results in quite acceptable mixing properties for
reasonably uncorrelated parameters (e.g., synchrotron and dust
spectral indices), other and more efficient methods may be required
for more complicated problems. Viable alternatives for such situations
are, e.g., rejection sampling or even standard Metropolis-Hastings
MCMC with proper burn-in monitoring. The details of the particular
sampling algorithm are of little importance as long as it can be
proved that the method produces samples from the correct conditional
distribution.

\section{Marginalization, priors and degeneracies}
\label{sec:degeneracies}

The algorithm described in \S\ref{sec:hybrid_sampling} provides
samples from the full joint posterior $P(\Bs, C_{\ell}, a_{\nu,i},
b_j, \mathbf{c}_k, \theta_k|\Bd)$.  From these multivariate
samples we estimate each parameter individually by marginalizing 
over all other parameters in the system and reporting, say, the marginal
posterior mean and standard deviation.

This is straightforward, but there are subtleties and care is required. 
Before applying the method to simulated data
in \S\ref{sec:verification} and \ref{sec:simulations}, therefore, we
discuss marginalization, priors, degeneracies, and high-dimensional probability
distributions.

\begin{figure*}[t]
\mbox{\epsfig{figure=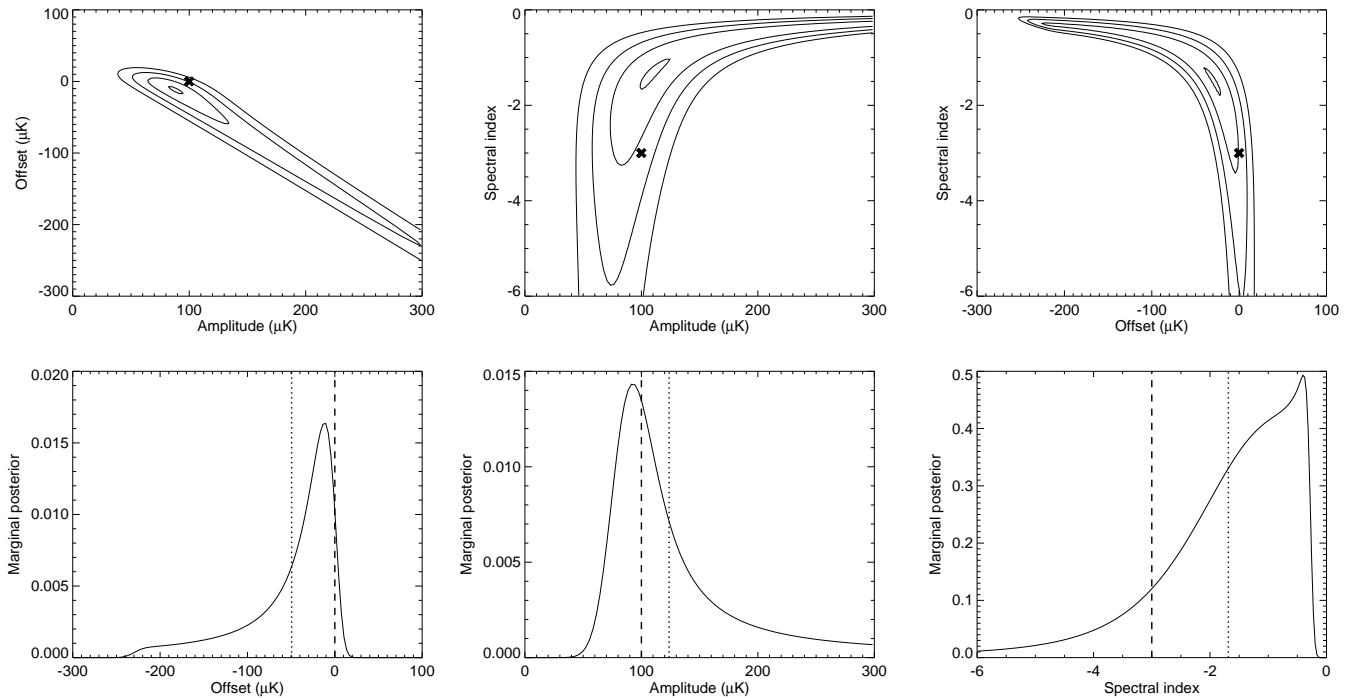,width=\linewidth,clip=}}
\caption{The one- (bottom row) and two-dimensional (top row) marginal
  posteriors for the single-pixel and four frequency-band data set
  described in \S\ref{sec:single_pixel}. The model includes a
  free offset for the lowest frequency, a foreground amplitude and a
  spectral index. The contours in the two-dimensional plots indicate
  where $-2\ln P$ has dropped by 0.1, 2.3, 6.17 and 11.8, respectively,
  corresponding to the peak and 1, 2, and 3$\sigma$ region for a
  Gaussian distribution. The crosses mark the true values. In the
  one-dimensional plots, the dashed lines indicate the true values,
  and the dotted lines show the marginal posterior mean.}
\label{fig:degenerate_case1}
\end{figure*}

Much of the following deals with the degeneracy between
unknown offsets (or monopoles) at each band and the overall zero-level
of a foreground component with a free amplitude at each pixel.
The same observations apply to any full-sky
template with a free amplitude at each band (e.g., the three
dipoles).  For simplicity we discuss only offsets below.  For the same reason, we
neglect the antenna-to-thermodynamic temperature conversion factor. 
When explicit formulae are
derived, the simplified and more readable versions are given in the
text;  full expressions are given in Appendices
\ref{sec:jeffreys_prior} and \ref{sec:full_constraint}.

It turns out that the degeneracy between unknown offsets and the
foreground zero-level has almost no effect on the CMB component.  
For the CMB, the relevant quantity is the sum over all
foregrounds, not internal degeneracies among different foregrounds.
If one cares only about separating the CMB from foregrounds, and not 
the foregrounds themselves, much of the following can be ignored.

\subsection{The offset-amplitude-spectral index degeneracy
  for a single pixel}
\label{sec:single_pixel}

Consider a hypothetical experiment that observes a single pixel at 30,
44, 70, and 100\,GHz, with RMS noise 10$\mu\textrm{K}$ in each band.
Assume that the signal is a straight power-law parametrized by
amplitude $A$ and spectral index $\beta$, and that the absolute offset
of the detectors is known perfectly for the three highest frequencies,
but not for the 30\,GHz band.  The signal model is
\begin{equation}
T_{\nu} = m \,\delta_{\nu,\nu_1} + A
\left(\frac{\nu}{\nu_1}\right)^\beta.
\label{eq:pix_model}
\end{equation}
There are three free parameters in this system, the
offset, amplitude, and spectral index, and four measurements.  Since
the number of constraints exceeds the number of degrees of freedom, it should  be
possible to estimate all three parameters individually.

We simulated one realization of this model, adopting the model
parameters $A = 100\,\muK$, $\beta = -3$ and $m=0\,\muK$, and adding
white noise to each band.  Our priors are chosen to be uniform 
over $-300\,\muK \le A, m \le 300\,\muK$ and $-6
\le \beta\le 0$.  We compute the joint posterior by a simple
$\chi^2$ evaluation over a $200\times200\times200$ grid, and
marginalize by direct integration.

Figure~\ref{fig:degenerate_case1} shows the results
in terms of one- and two-dimensional marginal posteriors. The true input values
are marked by thick crosses in the top panels and by dashed lines in
the bottom panels. The posterior means are shown by dotted lines in
the bottom panels.

This simple example highlights two problems that will recur in later
sections. First, as the top left panel shows, the offset and
amplitude are highly degenerate and anti-correlated; one may add
an arbitrary offset to the 30\,GHz band and subtract
it from the foreground amplitude, without affecting the
final $\chi^2$.  This degeneracy is a crucial issue for CMB component
separation.  Many foregrounds have power-law spectra, and differential
anisotropy experiments (e.g., WMAP) cannot determine absolute offsets. 
The monopoles of the WMAP temperature sky maps were
determined \emph{a posteriori} based on a co-secant fit to a
crude plane-parallel galaxy model \citep{bennett:2003b, hinshaw:2007}.
This approach is prone to severe modelling errors, precisely because
of this type of degeneracy.

The second problem is that integration over a highly degenerate joint posterior yields
complicated and strongly non-Gaussian marginal posteriors.  Obtaining
unbiased point estimates from these posteriors is not trivial.
Clearly, the posterior mean is not an unbiased estimator. 
Further, as we will see in the next section, even the posterior maximum is biased in
general, unless special care is taken when choosing priors.

\subsection{Uniform vs.~Jeffreys' prior}
\label{sec:flat_vs_jeffrey}

The strong degeneracies found in the previous example can be broken partially by adding
more data.  Consider a full-sky data set pixelized at HEALPix 
resolution $N_{\textrm{pix}} = 16$ (3072 independent pixels).  
Reduce the noise to $1\,\muK$ RMS per pixel.  Use the same signal 
model as before, but with an offset common to all pixels,
\begin{equation}
T_{\nu}(p) = m \,\delta_{\nu,\nu_1} + A(p)
\left(\frac{\nu}{\nu_1}\right)^{\beta(p)}.
\label{eq:offset_model}
\end{equation}
We adopt the spatially varying synchrotron model of \citet{giardino:2002}  
as a template for the amplitude and spectral index of the signal component 

We simulated a new data set, and computed the
marginal monopole posterior by direct integration. This is
straightforward because, for a given value of $m$, the conditional
amplitude-spectral index posterior reduces to a product of
single-pixel distributions. The integration therefore goes over a sum
of $N_{\textrm{pix}}$ two-dimensional grids, rather than a single
$2^{N_{\textrm{pix}}}$ grid.

\begin{figure}[t]
\mbox{\epsfig{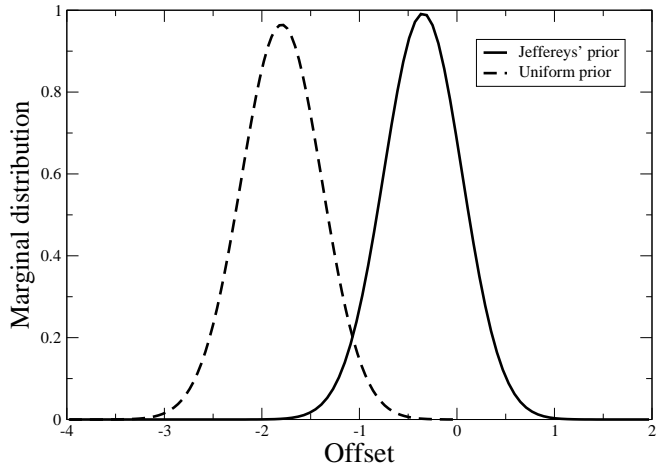}}
\caption{Comparison of the marginal offset posterior for a uniform
  (dashed line) and Jeffreys' ignorance (solid line) prior on the
  spectral index $\beta$. See \S\ref{sec:flat_vs_jeffrey} for a
  full discussion of both the model and details of the prior.}
\label{fig:jeffreys_vs_flat}
\end{figure}

The result is shown as a dashed curve in
Figure \ref{fig:jeffreys_vs_flat}.  Two points are noteworthy. 
First, the marginal distribution is nearly
Gaussian, in contrast to the strongly non-Gaussian single-pixel
posterior shown in the bottom panel of Figure \ref{fig:degenerate_case1}. 
Thus, the additional data seem to have
broken the degeneracy.  Second, however, the distribution has a 
mean and standard deviation of $-1.8\pm0.4$, 
more than 4$\sigma$ away from zero!  Repeated experiments with
different noise seeds gave similar results.

This behaviour is a result of the choice of prior.  We initially adopted a 
uniform prior on the offset, the amplitudes
and the spectral indices, with little thought to why we should
do so.  This was a poor choice. 
\citet{jeffreys:1961} argued that when nothing is known about a
particular parameter, one ought to adopt a prior that does not implicitly
prefer a given value over another, relative to the likelihood.
This is not in general the uniform prior.

Jeffreys argued that the appropriate
ignorance prior is given by the square root of the Fisher information
measure,
\begin{equation}
  P_{\textrm{J}}(\theta) \sim \sqrt{F_{\theta\theta}} = \sqrt{-\left<\frac{\partial^2 \ln \mathcal{L}}
  {\partial^2\theta}\right>},
\end{equation}
where the angle brackets indicate an ensemble average. This prior  
ensures that no parameter region is preferred based
on the parametrization of the likelihood alone; it is therefore a
proper ignorance prior \citep[e.g.,][]{box:1992}.

The log-likelihood corresponding to the model defined in Equation
\ref{eq:pix_model} reads
\begin{equation}
-2\ln \mathcal{L} = \sum_{\nu} \left(\frac{d_{\nu} - m \delta_{\nu,\nu_1} - A
  \left(\frac{\nu}{\nu_1}\right)^{\beta}}{\sigma_{\nu}}\right)^2.
\end{equation}
Computing the second derivatives of this expression with respect to
$A$, $m$, and $\beta$, we find that the appropriate Jeffreys' priors
for the three parameters are
\begin{align}
P_{\textrm{J}}(A) &\sim 1,\\
P_{\textrm{J}}(m) &\sim 1, \\
P_{\textrm{J}}(\beta) &\sim
\sqrt{\sum_{\nu}{\left(\frac{1}{\sigma_{\nu}} \left(\frac{\nu}{\nu_1}\right)^{\beta} \ln\left(\frac{\nu}{\nu_1}\right)\right)^2}},
\end{align}
respectively. In general, the ignorance prior for any linear parameter
in a Gaussian model is uniform because the second derivative
of the likelihood is constant. However, for non-linear parameters
greater care is warranted.

\begin{figure}[t]
\mbox{\epsfig{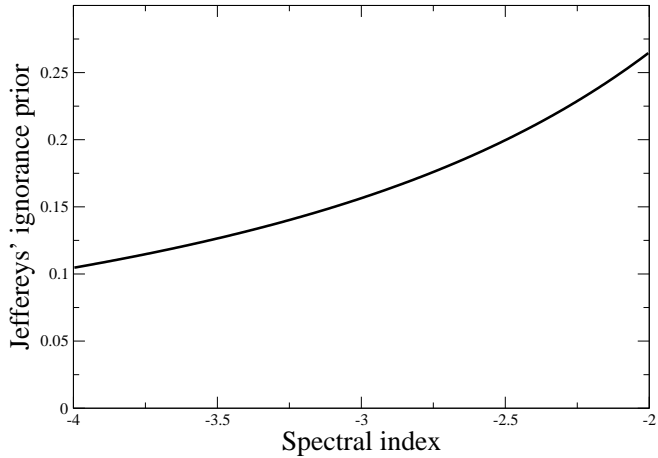}}
\caption{Jeffreys' ignorance prior for the spectral index $\beta$,
  defined by Equation \ref{eq:offset_model}. Steep indices
  are given less weight than shallow ones to
  compensate for their smaller overall impact on the likelihood.}
\label{fig:prior}
\end{figure}

Figure \ref{fig:prior} shows Jeffreys' prior for the spectral index
$\beta$, limited to $-4 \le \beta \le -2$. $\beta = -2$ is given about
two and half times more weight than $\beta = -4$.  Intuitively, this
is necessary because there is an asymmetry between a steep and a
shallow spectrum: A steep spectrum means that the signal dies off
quickly with frequency, while a shallow spectrum implies that it
maintains its strength longer. Thus, there is a larger allowed
parameter volume with steep indices than with shallow, leading to an
imbalance in terms of marginal probabilities. This parametrization
effect is countered by the Jeffreys' prior.

The solid curve in Figure \ref{fig:jeffreys_vs_flat} shows the result 
of using Jeffreys' prior instead of a uniform prior.  Similar behaviour is
observed independent of noise realization.  The conclusion is
clear: a proper ignorance prior leads to unbiased estimates,
while a naive uniform prior leads to biased estimates.

In addition to this basic ignorance prior, it may be beneficial to
adopt physical priors, based on knowledge from other experiments. For
example, if one had reason to expect that the dominant signal in a
given data set were Galactic synchrotron emission, a reasonable prior could be
$\beta = -3.0 \pm 0.3$, based on low frequency measurements. 
The physical prior is multiplied by the ignorance prior, taking account of
both effects.  In the rest of the paper, when we say that a Gaussian
prior is adopted for the spectral indices, we mean a
product of a Gaussian and Jeffreys' prior.

\subsection{Marginalization over high-dimensional and degenerate
  posteriors}
\label{sec:high_dimensional}

The previous section shows that given sufficient data and an
appropriate prior, the marginal posterior is a good estimator of the
target parameter.  In this section we investigate what happens 
 when the data are not sufficiently strong to break a
degeneracy.  We replace the single-channel offset $m$ by
a template amplitude $b$ coupled to a fixed free-free template
$t_{\textrm{ff}}(p)$ and a spectral index of $\beta_{\textrm{ff}} =
-2.15$,
\begin{equation}
  T_{\nu}(p) = b\, t_{\textrm{ff}}(p) \left(\frac{\nu}{\nu_1}\right)^{-2.15}  + A(p)
  \left(\frac{\nu}{\nu_1}\right)^{\beta(p)}. 
\end{equation}

Two modifications are made to the simulation. First, the spectral
index of the synchrotron component is fixed to
$\beta_{\textrm{s}}=-3$, rather than being spatially varying. Second,
a fifth frequency channel is added at 143 GHz. No free-free
component is added to the data; the optimal template
amplitude value is zero. The question is whether these
data are sufficient to distinguish between synchrotron and
free-free emission with similar spectral indices of $-3$ and
$-2.15$, respectively.

\begin{figure}[t]
 \mbox{\epsfig{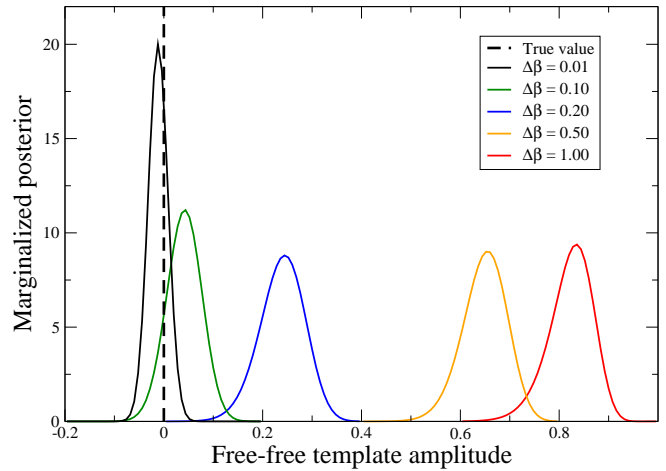}}
  \caption{Marginal free-free template amplitude posteriors for
    various priors on the synchrotron spectral index. See \S\ref{sec:high_dimensional} for a full discussion of this case.}
\label{fig:freefree_posterior}
\end{figure}

The answer is no.  Figure\,\ref{fig:freefree_posterior} shows the marginal template amplitude
posteriors, computed by direct integration as in the previous section.
The different curves correspond
to different Gaussian priors imposed on the synchrotron spectral
index.  All are centered on the true value $-3$, but with different standard deviations
$\Delta\beta_{\textrm{s}}$.  With the strong prior of
$\Delta \beta_{\textrm{s}} = 0.01$, the amplitude posterior is 
well-centered near the true value of zero.  However, when the prior is
gradually relaxed, the marginal posterior widens and drifts away
 from the true value.  The marginal posterior is not a useful estimator for the template
amplitude in this case.

This behaviour is explained by the fact that with 3072 independent
pixels the contribution of noise to the offset amplitude is insignificant
compared to the uncertainty introduced by coupling to the synchrotron
component.    Moreover, the amplitude and spectral index distributions are 
similar for the two foreground components.  As a result, the joint distribution
becomes long, narrow and curved, like that in  the top middle panel of 
Figure\,\ref{fig:degenerate_case1},  The marginal
one-dimensional posteriors are dominated by the ``boomerang wing''
orthogonal to the parameter axis. Similarly, the ``wing''  parallel to the axis is diluted. 
Given sufficiently strong degeneracies, the marginal distributions no longer contain the
maximum-likelihood point within their, say, $3\sigma$ confidence
regions.   When the prior is made increasingly tight, however, the wings of the
distribution are gradually cut off and the marginal distribution
homes in on the true value.  Thus, the collection of distributions
shown in Figure \ref{fig:freefree_posterior} in some sense visualizes
the joint posterior.

This behaviour may be quantified by means
of the covariance matrix of the Gaussian amplitude part of the system,
defined in Equation \ref{eq:coupling_matrix}.  A useful
quantity describing this matrix is its condition number, the ratio of
its largest and smallest eigenvalues.  For the particular case discussed
above, we find that the condition number is $4\times10^6$, which,
although tractable in terms of numerical precision for double
precision numbers\footnote{The absolute limit on the condition number
  for reliable matrix inversion is $10^{-6}$ for single-precision
  arithmetic and $10^{-12}$ for double precision.  However, in
  practice one should stay well below these values, in particular for
  iterative applications since small numerical errors may propagate in
  an uncontrolled manner.}, indicates a very strong degeneracy.

\subsection{The offset vs.\ amplitude degeneracy for full-sky data}
\label{sec:degen_fullsky}

The final example we consider before turning to realistic
simulations is the same as in \S\ref{sec:flat_vs_jeffrey}, except
we allow a free offset for \emph{all} frequency bands, not
just one.  (This is characteristic of real experiments, which do not know
the absolute zero-point at any frequency.)  The model is 
\begin{equation}
  T_{\nu} = m_{\nu} + A(p) \left(\frac{\nu}{\nu_1}\right)^{\beta(p)}.
\label{eq:offsets}
\end{equation}

If the spectral index is constant over the sky,
$\beta(p) = \beta$, this is a perfectly degenerate model:
\begin{align}
  T_{\nu} &= m_{\nu} + A(p) \left(\frac{\nu}{\nu_1}\right)^{\beta} \\
  &= (m_{\nu}+\delta m\left(\frac{\nu}{\nu_1}\right)^{\beta}) +
  (A(p)-\delta m) \left(\frac{\nu}{\nu_1}\right)^{\beta} \\
  &= m'_{\nu} + A'(p) \left(\frac{\nu}{\nu_1}\right)^{\beta}.
\label{eq:perfect_degeneracy}
\end{align}
One can simply add a constant to the foreground amplitude, and
subtract a correspondingly scaled value from each offset. 
It is thus impossible to determine individually the absolute
zero-level of foreground component and offsets.  To obtain physically
relevant results, external information must be imposed.

Spatial variations in the spectral index partially resolve this
degeneracy.  In the
\citet{giardino:2002} synchrotron model the spectral index
$\beta_{\textrm{G}}$ varies smoothly on the sky between $-2.5$
and $-3.2$.  The condition number of the foreground amplitude-offset
covariance matrix is $2\times10^7$, and the covariance matrix is 
no longer singular. A modified index model with ten times
smaller fluctuations but the same mean ($\beta(p) = 0.9
\left<\beta_{\textrm{G}}\right> + 0.1 \beta_{\textrm{G}}$) increases
the condition number by two orders of magnitude, to $2\times10^9$.

These strong degeneracies lead to the same quantitative behaviour as
seen for the marginal free-free template amplitude posterior in the
previous section, making it very difficult to estimate both all
offsets and the foreground amplitude zero-level individually. In
practice, external constraints are required. We have implemented two
approaches for dealing with this degeneracy in our code, both based on
the projection operator described in \S\ref{sec:amplitude_sampling}.

The first and more direct approach is to assume that the offsets of
one or more bands are known \emph{a-priori} by external information.
For instance, if an experiment somehow measured total power, as
opposed to differences alone, detailed knowledge about the instrument
itself could be used for these purposes. The advantage of this
approach is that it is exact, assuming the validity of the prior, and
the accuracy of all uncertainties is maintained. It is implemented
simply by demanding that $m_{\nu_0} = 0$, which requires, in terms of
the orthogonality vectors defined in \S\ref{sec:amplitude_sampling},
$q_{\nu,i} = \delta_{\nu,\nu_0} \delta_{i,0}$.  

The second approach is based on the observation that the degeneracy
between the foreground amplitude and the offsets seen in Equation
\ref{eq:perfect_degeneracy} leads to a very specific frequency
distribution of offset amplitudes. Specifically, $m'_{\nu} =
(m_{\nu}+\delta m(\nu/\nu_1)^{\beta})$, where $\delta m$ is an
arbitrary constant, \emph{but common to all frequency bands}. It is
therefore possible to require that the set of offsets should not have
a frequency spectrum that matches the foreground spectrum.

The corresponding constraint on $m_{\nu}$ may be derived from
\begin{equation}
\begin{split}
\chi^2 &= \sum_{\nu,pp'} \left(m_{\nu} - \delta
  m \left(\frac{\nu}{\nu_1}\right)^{\beta(p)}\right) N_{\nu,pp'}^{-1}\\&\quad\quad\quad\quad\quad\quad\quad\left(m_{\nu} - \delta
  m\left(\frac{\nu}{\nu_1}\right)^{\beta(p')}\right).
\end{split}
\end{equation}
by first taking the derivative with respect to $\delta m$, and then
enforcing a vanishing foreground component, $\delta m = 0$,
\begin{equation}
\sum_{\nu} m_{\nu} \left[\sum_{p,p'} N^{-1}_{\nu,pp'}
\left(\frac{\nu}{\nu_1}\right)^{\beta(p')}\right] = 0
\end{equation}
The expression in brackets says that the offsets should be orthogonal
to the mean noise-weighted foreground spectrum.

If the total signal model includes more than one signal component with
a free amplitude at each pixel, then these should be included jointly
in the above $\chi^2$. A particularly important case is that including
both a CMB signal, which has a frequency-independent spectrum, and a
proper foreground component. For this case, we have
\begin{equation}
\begin{split}
\chi^2 &= \sum_{\nu,pp'} \left(m_{\nu} - m_0 - \delta
  m \left(\frac{\nu}{\nu_1}\right)^{\beta(p)}\right)
N_{\nu,pp'}^{-1}\\&\quad\quad\quad\quad\quad\quad\quad\left(m_{\nu} -
  m_0 -\delta
  m\left(\frac{\nu}{\nu_1}\right)^{\beta(p')}\right),
\end{split}
\end{equation}
where $m_0$ is the additional degree of freedom introduced by the CMB
signal. The equivalent constraint on $m_{\nu}$ derived from this
expression is notationally more involved (see Appendix
\ref{sec:full_constraint} for a full derivation and constraints), but
may  be written as before in terms a set of
orthogonality vectors $q^i_{\nu}$.

While this orthogonality constraint is effective for estimating
the absolute zero-level of the foreground component in question, 
it corresponds to a strong implicit prior that is not likely to be compatible 
with reality.  If there are indeed random offsets at all frequencies, some
fractional combination of these offsets will mimic a foreground component. In the above
approach, this component is \emph{defined} to be a foreground signal,
rather than an offset. Further, no mixing between the two components
is allowed. Thus, the estimated error bars on both the offsets and
foreground zero-level will be underestimated.

Recall, however,  that this entire
discussion concerns the relative contributions to the foreground
zero-level and the free offsets, not the CMB signal, which relies
on the sum of the two components alone. The fact that the
\emph{estimated error} in the foreground zero-level is under-estimated
by a small factor, say, four or five ($\sigma_{\textrm{est}}
\sim0.5\,\muK$ vs.\ $\sigma_{\textrm{true}}\sim2\,\muK$; see the
simulation described in \S\ref{sec:simulations}), is of no
consequence for most applications. Far more important is the fact that
this approach provides excellent estimates of both the CMB sky signal
and the spectral index distribution, the two quantities where most of
the physics lie. This is in sharp contrast to the method employed by
the WMAP team, which is based on a co-secant fit to a plane-parallel
Galaxy model \citep{bennett:2003b, hinshaw:2007}. While that specific
approach is prone to severe modelling errors because of its lack of
detailed foreground modelling, the current approach is internally
consistent with respect to all signal components. For more discussion
on this issue, see Appendix \ref{sec:full_constraint}, as well as the
actual analysis of the 3-yr data presented by \citet{eriksen:2007c}.
In that analysis, a common offset of $\sim -13\,\muK$ is detected in all
frequency bands, as well as a significant residual dipole in the
V-band data.

\subsection{Summary}
\label{sec:summary}

\begin{figure}[t]
\mbox{\epsfig{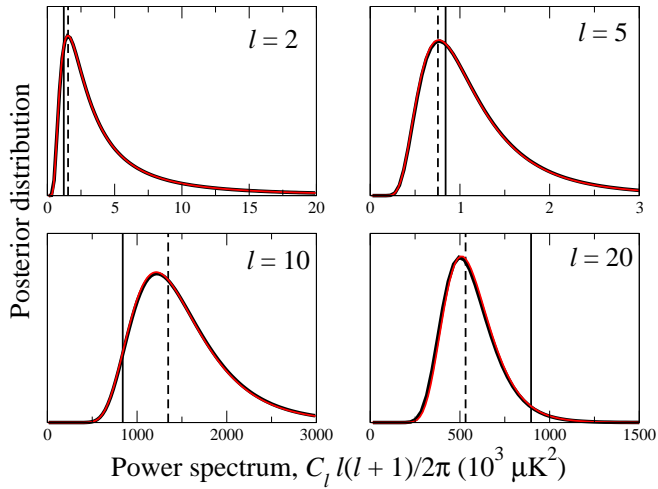}}
\caption{Verification of the CMB power spectrum distributions produced
  by Gibbs sampling. Black lines show analytically computed slices
  through the joint likelihood $\mathcal{L}(C_{\ell})$, and red lines
  shows the same computed by Gibbs sampling. Vertical lines show the
  theoretical input spectrum used to generate the simulation (solid
  lines) and the realization specific spectrum (dashed lines),
  respectively.}
\label{fig:verification_cmb}
\end{figure}

The above discussion may be summarized by the following observations:
\begin{itemize}
\item The marginal mean is a good estimator only for mildly degenerate
  and non-Gaussian joint distributions. Strongly degenerate models
  should be avoided, because they are difficult to summarize by simple
  statistics, and because it takes a prohibitive number of samples to
  fully explore them.
\item The uniform prior is a proper ignorance prior for Gaussian
  variables only. In general, Jeffreys' rule should be used in the
  absence of informative priors.
\item For experiments with unknown offsets at each frequency band,
  there is a strong degeneracy between these offsets and the
  overall zero-level of the foreground amplitudes. This degeneracy
  should be broken by external or internal priors, if marginal
  posteriors are to be used as estimators.
\end{itemize}

\section{Code verification}
\label{sec:verification}

In \S\,\ref{sec:degeneracies} we considered simple toy models to develop
 intuition about the target distributions. We used analytical, 
 brute-force computations to avoid the complexities of real-world computer code. 
 In this section, we turn our attention to Commander,  our
implementation of the joint foreground-CMB Gibbs sampler described
in \S\ref{sec:hybrid_sampling}.

Three conditional distributions are involved in this joint Gibbs
sampler, namely the CMB power spectrum distribution $P(C_{\ell}|\Bs)$,
the amplitude distribution $P(\Bs, a_{\nu,i}, b_j, \mathbf{c}_k |
C_{\ell}, \theta_k, \Bd)$, and the spectral parameter distribution
$P(\theta_k |\Bs, a_{\nu,i}, b_j, \mathbf{c}_k, \Bd)$. In the
following three subsections, we test the output from Commander for these
three conditional distributions against analytical expressions, at low 
resolution, to verify both the general sampling algorithms and our specific
implementation.

\subsection{The CMB power spectrum sampler}

To verify the CMB power spectrum distribution $P(C_{\ell}|\Bs)$, we
construct a low-resolution CMB-only simulation as follows.  Draw a
random CMB realization from a standard $\Lambda$CDM power spectrum
\citep{spergel:2007}, smooth to $10^{\circ}$ FWHM, and pixelize at
$N_{\textrm{side}}=16$.  Add white noise of $1\,\muK$ RMS to each
pixel.  Impose the WMAP Kp2 sky cut \citep{bennett:2003b}, without
point sources and downgraded to $N_{\textrm{pix}} = 16$, on the data.

We compute slices through the corresponding likelihood by
considering each $\ell$ individually, fixing all other
multipoles at the input power spectrum, with a brute-force calculation in pixel-space
\citep[e.g.,][]{eriksen:2007a},  and with Commander.  The outputs
from the latter are smoothed through Rao-Blackwellization
\citep{chu:2005} to reduce Monte Carlo errors.

Figure\,\ref{fig:verification_cmb} shows the results for four multipoles. 
The theoretical input spectrum is shown by vertical solid lines, and the true realization
spectrum by dashed lines.   Commander reproduces the CMB power spectrum
distributions perfectly.

\subsection{The Gaussian amplitude sampler}

\begin{figure}[t]
\mbox{\epsfig{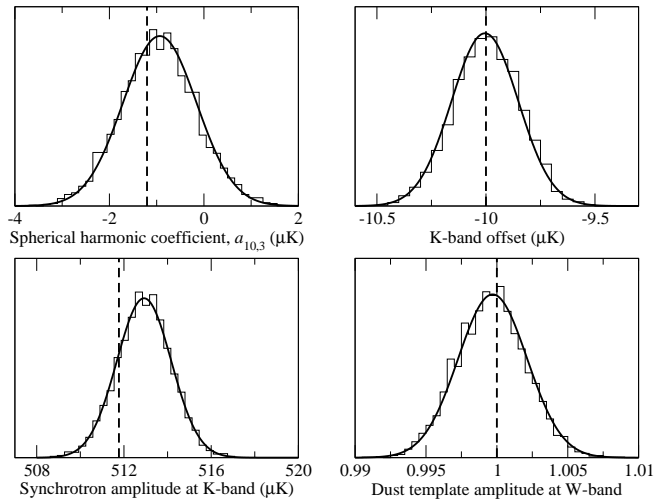}}
\caption{Verification of Gaussian amplitude sampler. Analytical
  marginal posteriors are shown by smooth distributions, and results
  from Commander are shown by histograms. The true values are
  indicated by vertical dashed lines.}
\label{fig:verification_gauss}
\end{figure}

\begin{figure}[t]
\mbox{\epsfig{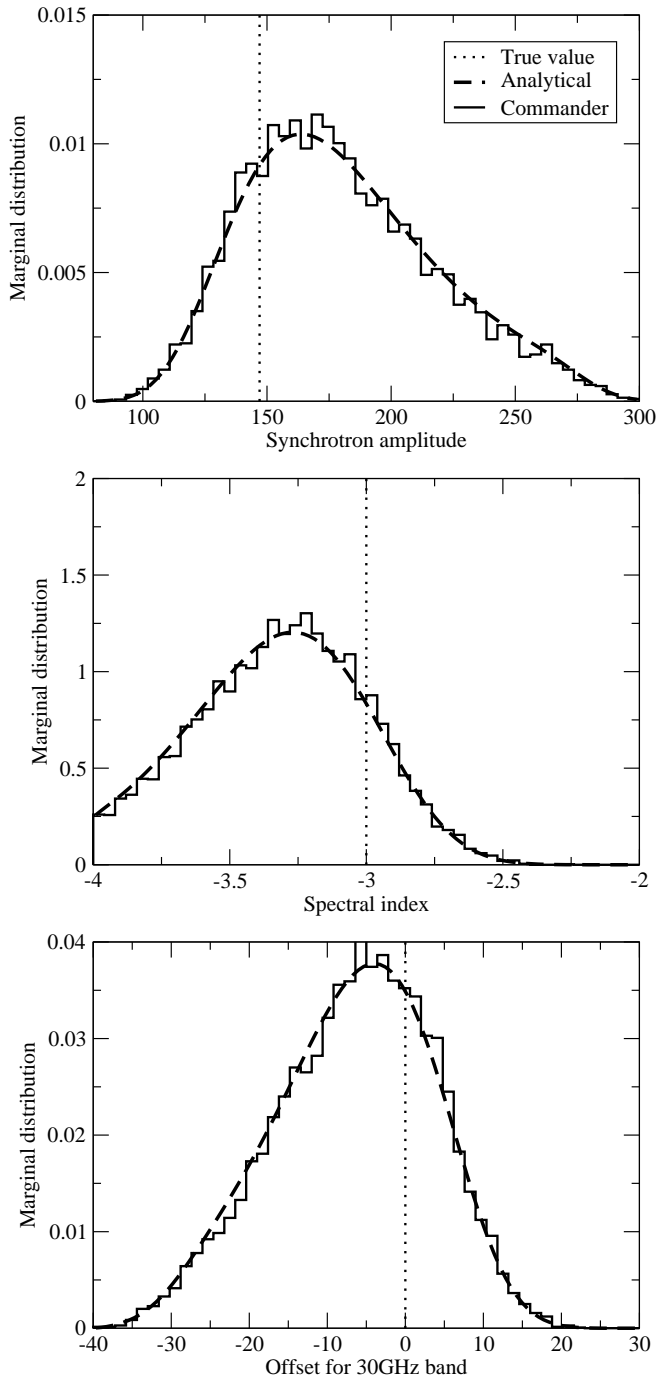}}
\caption{Comparison of marginal posteriors for a single pixel,
  computed both analytically (thick, dashed curve) and with
  Commander (thin line). The true value is indicated by a vertical
  dotted line.}
\label{fig:verification_single_pixel}
\end{figure}

To verify the amplitude distribution $P(\Bs, a_{\nu,i}, b_j,
\mathbf{c}_k | C_{\ell}, \theta_k, \Bd)$, we construct a simulation at
$N_{\textrm{side}}=8$ (768 independent pixels, angular resolution
$20^{\circ}$ FWHM). The CMB realization is the same as in the previous
section, appropriately smoothed.  Five frequency channels are
simulated, corresponding to the five WMAP channels.  In addition to
the CMB sky signal, $\Bs$, we add a synchrotron signal, $\Bc(p)$ with
a spatially varying spectral index, a dust template with an amplitude,
$b$, scaled to unity at W-band, and a $a_{0} = -10\,\muK$ monopole to
the K-band. (See foreground description in Section \ref{sec:model} for
further details on this model.) Thus, all four types of amplitudes are
represented. White noise of $1\,\muK$ RMS is added to each pixel at
each frequency.

We fix the CMB power spectrum and synchrotron spectral index map,
and compute the joint Gaussian amplitude distribution both
analytically and with Commander. The analytical
computation is performed by direct evaluation of the mean $\hat{\Bx}$
and covariance matrix $\mathcal{A}$ defined by 
Equations~\ref{eq:coupling_matrix} and \ref{eq:joint_mean}. The marginal
variances of each parameter are given by the diagonal elements of
$\mathcal{A}$.

Figure \ref{fig:verification_gauss} shows the marginal distributions
for one parameter of each type.  Again, Commander reproduces the 
exact analytical result perfectly.

\subsection{The spectral index sampler}

To verify the spectral index sampler for $P(\theta_k |\Bs, a_{\nu,i},
b_j, \mathbf{c}_k, \Bd)$, a single-pixel distribution, we simulate a
single pixel.  The signal model is identical to that in
\S\ref{sec:single_pixel}, comprising a synchrotron component with
unknown amplitude and spectral index, plus an unknown offset at the
lowest frequency.

We compute the corresponding three-dimensional joint posterior by
direct grid evaluation and by Commander.
Figure~\ref{fig:verification_single_pixel} shows the corresponding
marginal distributions.  Again we find perfect agreement.

All conditional distributions currently implemented in
Commander have thus been verified.

\section{Application to simulated 3-yr WMAP data}
\label{sec:simulations}

We turn now to a more realistic simulation, with properties
corresponding to the 3-yr WMAP data. The simulation has two goals.
First, to show that the method can handle data with realistic
complexities, and that it is applicable to the current WMAP data and
(even more importantly) the up-coming Planck data. Second, to provide
the necessary background for understanding the results from the actual
3-yr WMAP analysis presented by \citet{eriksen:2007c}. 

\subsection{Simulation, model and priors}
\label{sec:model}

We construct the simulation as follows. Draw a CMB sky realization
from the best-fit $\Lambda$CDM power spectrum presented by
\citet{spergel:2007}.  Convolve with the each of the beams of the ten
differencing assemblies of WMAP \citep{bennett:2003a}, pixelized at a
HEALPix resolution of $N_{\textrm{side}}=512$.  Add white noise to
each map with standard deviation $\sigma(p) = \sigma_0 /
\sqrt{N_{\textrm{obs}}(p)}$, where $N_{\textrm{pix}}$ is the number of
observations at pixel $p$ (provided on
Lambda\footnote{http://lambda.gsfc.nasa.gov} together with the actual
sky maps).

Downgrade these ten maps to a common resolution of $3^{\circ}$ FWHM
and $N_{\textrm{pix}} = 64$, bandlimiting each map at
$\ell_{\textrm{max}} = 150$. Create frequency maps by co-adding
differencing assembly maps at the same frequency (e.g.,
Q~=~(Q1+Q2)/2). Add uniform white noise of $2\,\muK$ RMS to each
frequency map to regularize the noise covariance matrix.

\begin{figure}[t]
\mbox{\epsfig{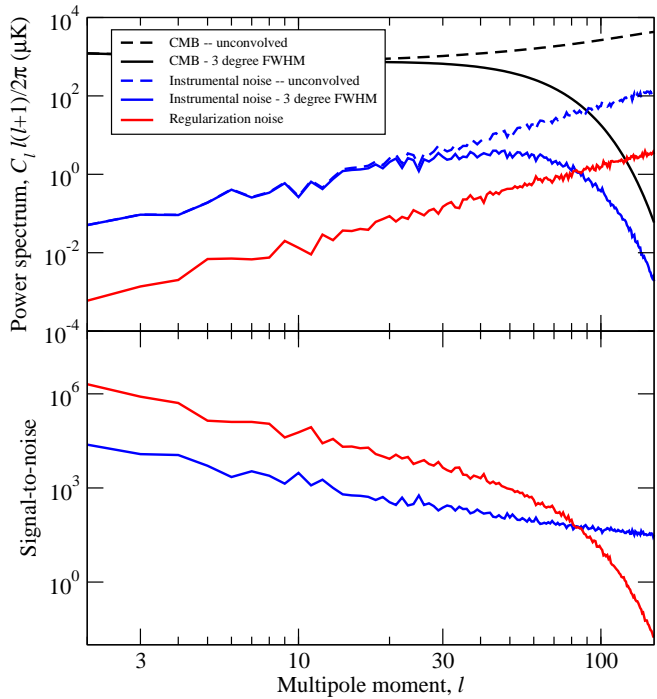}}
\caption{Top panel: CMB (black lines) and noise power spectra (blue
  and red lines) for the 3-yr WMAP V-band channel. Solid lines show
  spectra for the $3^{\circ}$ FWHM data set, and dashed lines show
  spectra at the native resolution of the V-band data. Bottom panel:
  CMB signal to noise ratio for instrumental (blue) and regularization
  (red) noise, respectively.}
\label{fig:signal_to_noise}
\end{figure}

Figure \ref{fig:signal_to_noise} shows the CMB and noise spectra of
the co-added V-band data, both at the native resolution of the
frequency band (dashed lines) and at the common $3^{\circ}$ FWHM
resolution (solid lines). The CMB to regularization noise ratio is
unity at $\ell \sim 120$, and less than 2\% at $\ell_{\textrm{max}} =
150$. Both the instrumental and the regularization signal-to-noise
ratios are $\gtrsim 500$ at $\ell \le 50$, and therefore negligible at
these scales compared to cosmic variance.

Instrumental noise averaged over the full sky is larger than the
regularization noise everywhere below $\ell \lesssim 80$. In the
ecliptic plane, where the instrumental RMS is about a factor of two
larger than the full-sky average due to WMAP's scanning strategy, it
dominates below $\ell \lesssim 100$. The result of this unmodelled
noise term is, as we will see later, a somewhat high pixel-by-pixel
$\chi^2$ in the ecliptic plane. However, since this error term is
correlated only on very small scales (the beam size of $3^{\circ}$
FWHM), and we understand its origin and benign behaviour, it does not
represent a significant problem for the analysis.  With additional
years of WMAP observations, and the addition of the Planck data, this
noise contribution will be further suppressed. Further, we will
consider in the future various approaches for taking this term into
account, for instance by computing explicitly the corresponding sparse
covariance matrix.

Our foreground model has three components, synchrotron, free-free, and
thermal dust emission.  For synchrotron emission, where the spectral
index is known to vary substantially with position on the sky, we
extrapolate the 408 MHz map \citep{haslam:1982} using a map of the
spectral index for each pixel. For the latter we use an updated
version of the Giardino et al. (2002) spectral index map that is based
on 408 MHz and WMAP 23GHz data, after removing the free-free emission
via the WMAP MEM free-freemodel (Bennett et al. 2003b)\footnote{These
  models were produced as part of the development of the Planck Sky
  Model, under the coordination of Planck Working Group 2.}.  The
free-free model is defined by the $H_{\alpha}$ template of
\citet{finkbeiner:2003}, scaled to 23\,GHz assuming an electron
temperature of $T_{\textrm{e}} = 4000\,\textrm{K}$ and a spatially
constant spectral index of $\beta_{\textrm{ff}} = -2.15$.  The dust
model is based on model 8 of \citet{finkbeiner:1999}, evaluated at
94\,GHz and scaled to other frequencies using a single-component
modified black-body spectrum with $T_{\textrm{d}} = 18.1\,\textrm{K}$
and an emissivity index of $\alpha = 2.0$. Anomalous dust is ignored
in this analysis.

Guided by the results of \citet{eriksen:2007c}, we add a common offset
to all frequencies of $-13\,\muK$.  No dipole contributions are added
to the simulations.

For the power spectrum analysis in \S\ref{sec:cosmology}, we analyze
the same realization with and without foregrounds, but with the same
sky cut. This allows us to distinguish between sky-cut and
foreground-induced effects.

We adopt the same parametric signal model as that used by
\citet{eriksen:2007c},
\begin{equation}
\begin{split}
  T_{\nu}(p) &= s(p) + m^{0}_{\nu} + \sum_{i=1}^{3} m^i_{\nu}
  \left[\hat{\mathbf{e}}_i\cdot\hat{\mathbf{n}}(p)\right] + \\+
  b&\left[t(p) a(\nu) \left(\frac{\nu}{\nu_0^{\textrm{dust}}}\right)^{1.7}\right] + f(p) a(\nu)
  \left(\frac{\nu}{\nu_0}\right)^{\beta(p)}.
\end{split}
\end{equation}
The first term is the CMB sky signal. The second and third terms are
the monopole $m^{0}$ and three dipole components $m^i$ defined by
standard Cartesian basis vectors. The fourth term is a dust tracer,
based on the FDS template coupled to a fixed spectral index of
$\beta_{\textrm{d}}=1.7$, and a free overall amplitude $b$. The
postulated power-law spectrum does not match the modified black-body
spectrum used to create the simulation, and modelling errors are
therefore to be expected. The fifth term is a single low-frequency
foreground component with a free amplitude $f(p)$ and spectral index
$\beta(p)$ at each pixel $p$. The antenna-to-thermodynamic
differential temperature conversion factor is $a(\nu)$, as always.

In addition to the previously described Jeffreys' prior, we adopt a prior of $\beta =
-3\pm0.3$ for the low-frequency foreground spectral index, assuming
that the foreground signal is synchrotron emission unless the data
require otherwise. This is not a particularly strong prior:
The free-free spectral index of $\beta_{\textrm{ff}}=-2.15$ is only
2.8$\sigma$ away from the prior mean, and it does not take a large
free-free amplitude to overcome this.  For instance, near the galactic
plane the standard deviation of the marginal index posterior is
$\sim0.01$,  thirty times smaller than the prior width.  At high latitudes, on 
the other hand,  the synchrotron spectral index is for
all practical purposes unconstrained.   The prior prevents this component 
from interfering with the CMB signal in regions where its amplitude is low.

We impose the orthogonality constraint discussed in \S\ref{sec:degen_fullsky}
to break the degeneracy between the free monopoles and dipoles 
at each band, and the foreground zero-level
and dipole.  An important goal in the following is to see whether this
approach yields sensible results.

With the simulation, model, and priors defined, we compute the joint
and marginal posteriors using the machinery described earlier in the
paper. The wall-clock time for generating one single sample is $\sim
50$\,s, parallelized over five 2.6\,GHz AMD Opteron 2218 processors,
one for each frequency band.  We generate five chains with 1000 samples
each, for a total wall clock time of 14\,hr. The total computational
cost is 350\,CPU hours.

\subsection{Burn-in, correlation lengths and convergence}

\begin{figure}[t]
\mbox{\epsfig{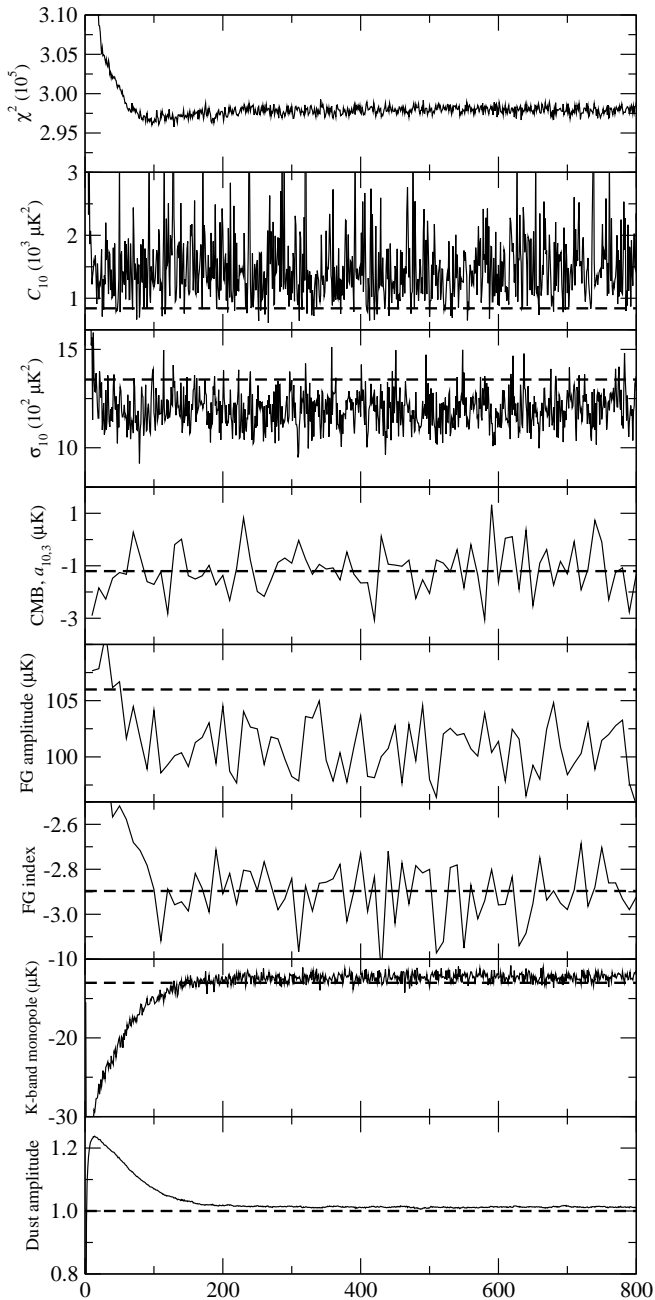}}
\caption{Trace plots showing the evolution of the Gibbs chains as a
  function of iteration count, for each type of parameter. The true
  input values are indicated by a horizontal dashed line, where
  applicable. The sky map type parameters are thinned by a factor of
  ten to reduce disk space requirements.}
\label{fig:wmap_trace_plots}
\end{figure}

We begin our examination of the results by plotting the output Markov
chains as a function of iteration count in Figure
\ref{fig:wmap_trace_plots}.  Each panel shows the evolution of one
parameter, such as the CMB power spectrum coefficient for a
single multipole or the dust template amplitude.

Burn-in is a crucial issue for Markov chain algorithms.  The chains
were initialized with a random CMB power spectrum over-dispersed
relative to the true distribution, and the spectral indices of the
low-frequency foreground component were drawn randomly and uniformly
between $-4$ and $-2$.  The Gibbs sampler needs some time to converge
to the equilibrium distribution; as we see in
Figure~\ref{fig:wmap_trace_plots}, about 200 iterations are required
to reach the equilibrium state.

\begin{figure*}[t]
\mbox{\epsfig{figure=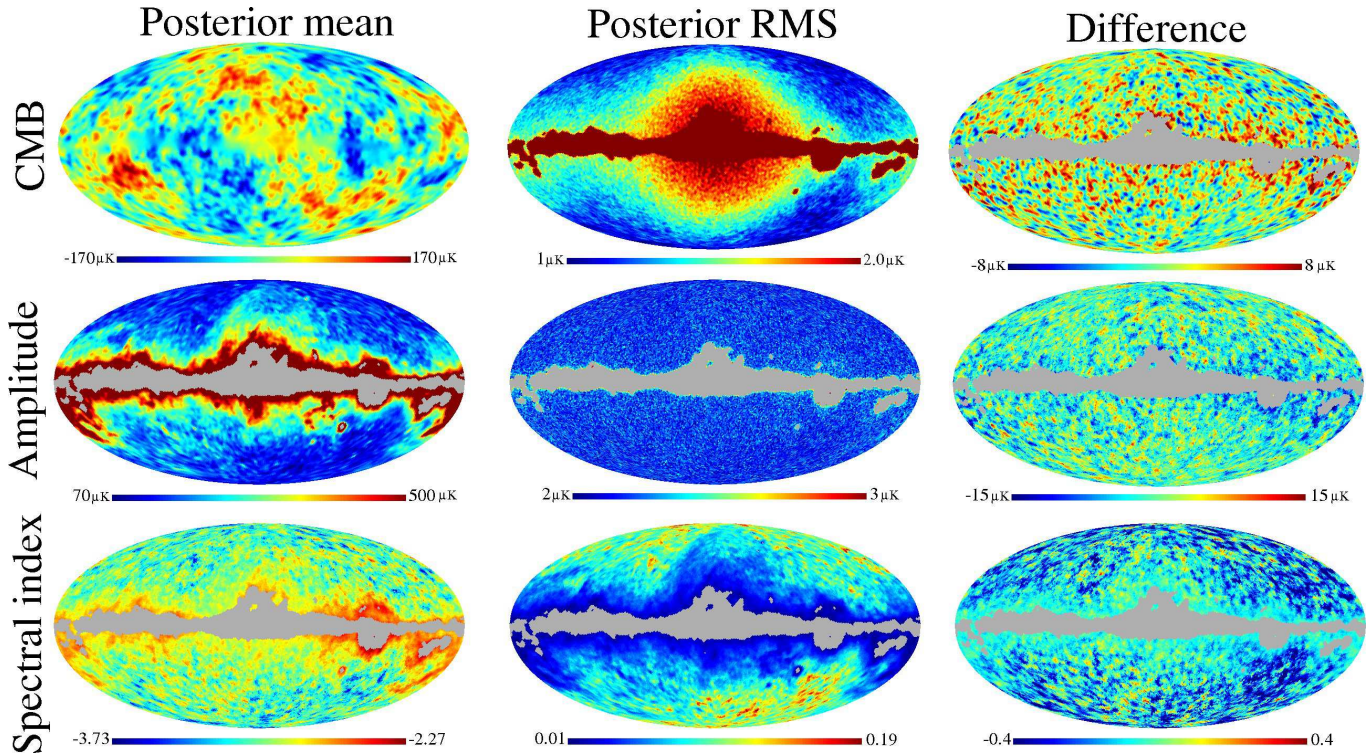,width=\linewidth,clip=}}
\caption{Marginal posterior sky signals from the WMAP simulation.  The
  left column shows the posterior mean for each pixel, the middle
  column shows the posterior standard deviation, and the right column
  shows the difference between the estimated posterior mean and the
  known input signal. From top to bottom, the rows show 1) the CMB
  solution; 2) the low-frequency foreground amplitude solution; and 3)
  the low-frequency spectral index solution. }
\label{fig:wmap_maps}
\end{figure*}

The last parameters to equilibrate are the global
monopole and dust amplitudes, because the uncertainty in these
very high signal-to-noise parameters is very small, and only small
steps can be made between consecutive Gibbs samples. Moreover, since 
these are global parameters, they couple to all other parameters.

The $\chi^2$ trace plot shows an interesting feature.  After
reaching a minimum $\chi^2$ solution after about 100~iterations, the
chain stabilizes at a very slightly higher equilibrium value.
This is due to the fact that the full distribution consists not only
of the sky signal components, but also the CMB power spectrum.
Maximizing the total joint posterior value is therefore a compromise
between minimizing the sky signal $\chi^2$ and optimizing the CMB
power spectrum posterior. At iteration number 100, the CMB component
is still burning in, whereas the foreground amplitude, the single most
important parameter in terms of $\chi^2$, has already reached its
equilibrium. The Markov chain thus overshoots in $\chi^2$
minimization until the CMB power spectrum equilibrates.

Correlation length is a second crucial issue for Markov chain algorithms. 
In general, classic Metropolis-Hastings 
algorithms have a long correlation length because they propose
relatively small modifications at each iteration in order to maintain
high acceptance probability. The Gibbs sampler works differently.
Because it samples from exact conditional distributions, large jumps
are perfectly feasible, at least in the absence of strong conditional
correlations.  (In the present case there are no such strong
correlations.)  The CMB power spectrum and CMB sky signal are only
weakly correlated in the high signal-to-noise regime, and the
foreground spectral index couples only moderately strongly to the
foreground amplitude of the same pixel, and weakly to anything else.
The result is excellent mixing properties and short correlation
lengths.

This translates into a high sampling efficiency and a relatively
small number of samples required for convergence. To quantify this, we
adopt the widely used Gelman-Rubin $R$ statistic \citep{gelman:1992},
which is the ratio between two variance estimates. If the Markov
chains have converged, the two estimates should agree, and their
ratio, $R$, should be close to unity. A typical recommendation is that
$R$ should be less than 1.1 to claim convergence, given that the
chains were initially over-dispersed, although smaller numbers are
clearly better.

Computing this statistic for the five chains above, while discarding
the first 200 samples, we find that $R$ is less than 1.01 for the CMB
power spectrum up to $\ell \sim 100$, less than 1.05 for the both the
CMB pixel and foreground amplitudes all over the sky, and less than
1.01 for the template amplitudes. Thus, even with such a relatively
modest number as 4000 samples, excellent convergence has been reached
on all marginal statistics. We return to the question of joint
convergence of the CMB spectrum posterior in \S\ref{sec:cosmology}.

\subsection{Component separation results}

We now turn to the marginal distributions of the
estimated signal parameters, and focus first on the signal components.
The CMB power spectrum is discussed separately in the next section.
The sky map results are summarized in Figure \ref{fig:wmap_maps} in
terms of the marginal posterior means, standard deviations, and
differences between the posterior means and the input maps. 
Table~\ref{tab:monopole} lists the monopole and dipole results. The dust
template amplitude posterior mean and standard deviation is $b =
1.013\pm0.002$.

Considering first the left column in Figure \ref{fig:wmap_maps}, we
see that the three sky map reconstructions are visually compelling. 
No obvious foreground residuals are observed in the CMB
map, familiar structures such as the North Galactic Spur and Gum
Nebula are seen in the foreground amplitude map, and the spectral
index map distinguishes clearly between the known synchrotron and
free-free regions.

\begin{figure}[t]
\mbox{\epsfig{figure=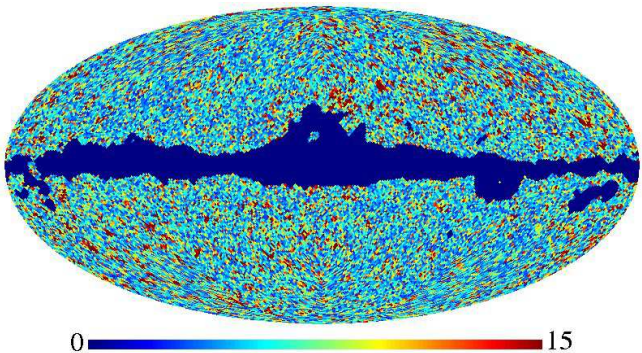,width=\linewidth,clip=}}
\caption{Mean $\chi^2$ map computed over posterior samples. A value of
  $\chi^2=15$ corresponds to a $\chi^2$ that is high at the 99\%
  significance level.}
\label{fig:wmap_chisq}
\end{figure}

These visual considerations are quantified  in the
right column, where the input maps has been subtracted from the
posterior means\footnote{For the foreground amplitude, the input map
  was estimated by fitting a single power law to the sum of the
  synchrotron and free-free components.}.  We see that the CMB
map has residuals at the $\sim3\,\muK$ RMS level, with a peak-to-peak
amplitude of $\pm8\,\muK$.  Little of these residuals is 
correlated on the sky except for a few patches near the galactic
plane. Most of the differences are simply due to instrumental noise.

For the foreground amplitude, more distinct correlated patches are
seen, in particular in regions with strong free-free emission. This is
due to the fact that a single power law is not a sufficiently good
approximation to the sum of the free-free and synchrotron components,
relative to the statistical uncertainty.

\begin{deluxetable}{lcccc}
\tablewidth{0pt} 
\tabletypesize{\small} 
\tablecaption{Monopole and dipole posterior statistics\label{tab:monopole}}
\tablecolumns{5}
\tablehead{ & Monopole & Dipole X & Dipole Y & Dipole Z \\
Band & ($\mu\textrm{K}$) & ($\mu\textrm{K}$)& ($\mu\textrm{K}$)&
 ($\mu\textrm{K}$)
}
\startdata

K-band    & $-12.4 \pm 0.5$ & $-0.6 \pm 0.7$ & $-0.4 \pm 0.5$ & $-0.1  \pm 0.1$ \\
Ka-band   & $-10.3 \pm 0.5$ & $-1.2 \pm 0.7$ & $-0.7 \pm 0.5$ & $-0.2 \pm 0.1$ \\
Q-band    & $-10.8 \pm 0.5$ & $-1.1\pm 0.7$  & $-0.7 \pm 0.5$ & $-0.1  \pm 0.1$\\ 
V-band    & $-12.3 \pm 0.5$ & $-0.6\pm 0.7$ & $-0.6 \pm 0.5$ & $ -0.1  \pm 0.1$ \\ 
W-band    & $-12.8 \pm 0.5$ & $-0.3\pm 0.7$ & $-0.2 \pm 0.5$ & $\,\,\,-0.1  \pm 0.1$

\enddata
\tablecomments{Means and standard deviations of the marginal monopole and dipole posteriors.}
\end{deluxetable}

Finally, even the the spectral index difference map shows clearly
correlated regions, and additionally a negative bias of about $-0.1$.
This bias is primarily due to two effects. First, as reported at the
beginning of this section, the dust template amplitude is
over-estimated by 1--2\%, mainly because of mis-specification of the
dust spectrum. As a result, slightly too much signal is subtracted
from the higher frequency channels, and this in turn steepens the
spectral index of the remaining signal. Second, at high latitudes the
data are noise dominated, and the $\beta = -3\pm0.3$ prior becomes
active. Because the true signal has an average of $\sim -2.9$ at high
latitudes, a bias of $\sim -0.1$ results.

Comparing the actual difference maps with the estimated errors shown
in the middle column of Figure \ref{fig:wmap_maps}, we see that the
errors of the CMB and foreground amplitudes are
underestimated by a factor of $\sim 1.5$ to 2. (These plots are
typically scaled to a dynamical range of $\sim\pm3\sigma$. The
expected peak-to-peak range in a difference plot is therefore roughly
three times the RMS error.) This is due to modelling errors in two
forms. First and foremost, we neglected the smoothed instrumental
noise in our data model, and this causes a significant unmodelled
uncertainty at the smoothing scale. However, being random with zero
mean, it does not induce significant structure on larger scales, and
it therefore has negligible impact on the scales of cosmological
interest ($\ell \le 30$). Second, the foreground model is simplified
compared to the input, as we approximate the sum of two power law
components by a single power law, and also assume a simple power-law
dust spectrum while the input sky has a modified black-body spectrum.
Combined, these effects introduce errors not captured by the estimated
statistical uncertainties.

Table~\ref{tab:monopole} lists the posterior mean and standard
deviations of the monopole and dipole coefficients. Recall that the
input parameters in the two cases were $-13\,\muK$ and $0\,\muK$,
respectively. In general, these values are reconstructed reasonably
well, although the error estimates are somewhat underestimated for
the Ka- and Q-band monopoles.  We see that the
orthogonality constraint described in \S\ref{sec:degen_fullsky}
is quite effective, producing a good estimate of all
quantities of interest. On the other hand, it does have the effect of
artificially reducing the error bars on the template amplitudes
somewhat, and also correlating them. However, misestimation of
the monopole error estimates by a few microkelvin is a small price to pay
for an absolute estimate of the foreground amplitudes to within a few
percent.

The features seen in the CMB RMS map may be understood qualitatively
in terms of the above results. First, the most dominating structure is
a hot-spot centered on the Galactic plane. This is mainly due to the
coupling between the galactic foregrounds and the x-component of the
dipole. Because of the large foreground signal in this direction, it
is hard to estimate the corresponding dipole component (see Table
\ref{tab:monopole}), and this transfers uncertainty from low to high
latitudes. Note, however, that this particular component has a very
specific correlation structure on the sky, which is taken implicitly
into account by the algorithm; this uncertainty does therefore not
significantly affect high-$\ell$ modes in the power spectrum, even
though it looks visually dominating in a marginal RMS map. Second, the
masked Galactic plane has a very high uncertainty, although not
infinite; the requirement of isotropy implies that the modes inside
this plane is to some extent restricted, at least on large angular
scales. Finally, as expected there is a (weaker) correlation between
the foreground amplitude and spectral index maps and the CMB RMS map.

Figure~\ref{fig:wmap_chisq} shows the average $\chi^2$ computed over
the 4000 accepted samples. A value of 15 in this plot corresponds to a
model that is excluded at the 99\% confidence level. Two points are
worth noticing in this plot. First, the ecliptic plane stands out with
higher $\chi^2$ values. As described above, this is due to the
unmodelled, smoothed instrumental noise. At a smoothing scale of
$3^{\circ}$ FWHM, this component is not fully negligible for the 3-yr
WMAP data relative to the CMB signal, and therefore causes a slight
bias at the smallest scales, $\ell \gtrsim 100$. However, we are
mainly interested in $\ell\le50$, and in this range the instrumental
signal-to-noise ratio exceeds 100 everywhere. This term does not
affect the CMB signal of primary interest.

The actual foreground-induced modelling errors are very small.
Indeed, despite the fact that we approximate the sum of two different
power laws with a single component, and assume an incorrect dust
spectrum, the $\chi^2$ distribution is essentially perfect near the
ecliptic poles, and the residuals are very small even close to the
Galactic plane.

However, the $\chi^2$ for the global solution as a whole is somewhat
poor, with a reduced $\chi^2$ of 1.68. This large value is largely
dominated by unmodelled smoothed instrumental noise in the ecliptic
plane, as discussed above; when analyzing the same data set at a
smoothing scale of $6^{\circ}$, rather than $3^{\circ}$, we found a
reduced $\chi^2 \sim 1.1$. Thus, when using the products from this
analysis in subsequent studies (e.g., for cosmological parameter
estimation), it is important to include only those scales that are
unaffected by the degradation process itself."

In summary,  the overall results are very promising indeed. 
The CMB sky signal is reconstructed to within a few percent
everywhere, as is the foreground amplitude. Further, the spectral
indices are accurate to the $\sim0.1$ level wherever there is a
significant signal, and the monopole and dipole coefficients are very
close to the true values. Finally, even the reconstructed dust
template amplitude is correct to within 2\%. 

\begin{figure}[t]
\mbox{\epsfig{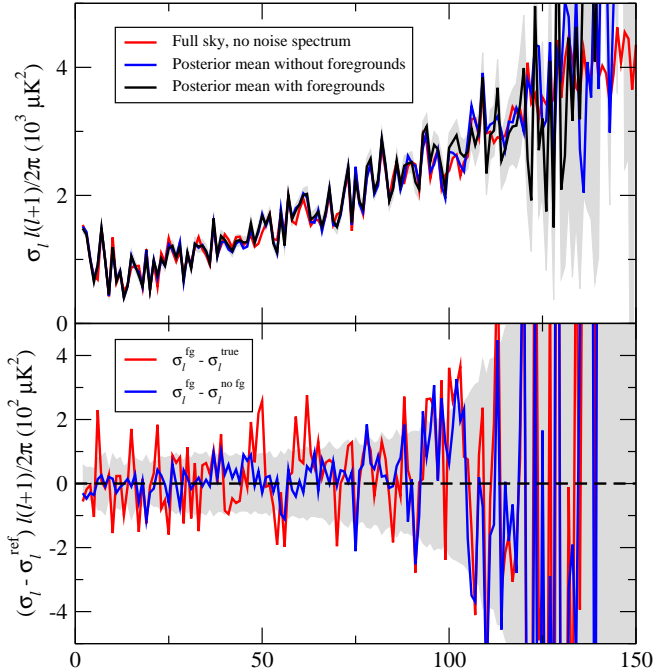}}
\caption{Top panel: Comparison of the posterior mean realization power
  spectra, $\sigma_{\ell}$, 1) as computed on the full sky (i.e., true
  realization spectrum; red line); 2) as computed on the cut sky, but
  not including foregrounds either in the simulation or in the model
  (blue line); and 3) as computed with Commander on the full
  simulation, including the foreground complexity (black line).
  The gray area indicates the $1\sigma$ confidence region for the case
  that includes foregrounds. Bottom panel: Difference between the
  spectra shown above; blue line shows the difference with and without
  foregrounds, and the red line shows the difference between the
  spectra including foregrounds and a sky cut and the true spectrum .}
\label{fig:wmap_sigma}
\end{figure}

The results are slightly more mixed when it comes to estimation of
uncertainties. For components with a relatively large intrinsic
uncertainty, such as the CMB sky signal and foreground spectral index,
the error estimates are quite reasonable. On the other hand, for
parameters with a high intrinsic signal-to-noise ratio, most
noticeably the dust template amplitude, the errors are clearly
under-estimated because of significant modelling errors. (We note that
analysis of simulations with a foreground composition and priors that
matches the assumed model yields, as expected from the results shown
in Section \ref{sec:verification}, both point estimates and
uncertainties in agreement with expectations.)

The remaining and key issue is what the impact of these residuals and
increased uncertainties are on the CMB power spectrum and cosmological
parameters at $\ell \le 30$. This is the topic of the next section.

\subsection{CMB power spectrum and cosmological parameters}
\label{sec:cosmology}

We now consider the CMB power spectrum posterior with the goal of
understanding the impact of both residual foregrounds and error
propagation on the final results. To do so, we consider both the identical 
simulation described in the previous section and a similar
one in terms of instrumental properties and sky coverage, but
excluding foregrounds both from the simulated data and the model.
Comparing the two against each other allows us to disentangle the
effects of foregrounds and sky cut.

\begin{figure}[t]
\mbox{\epsfig{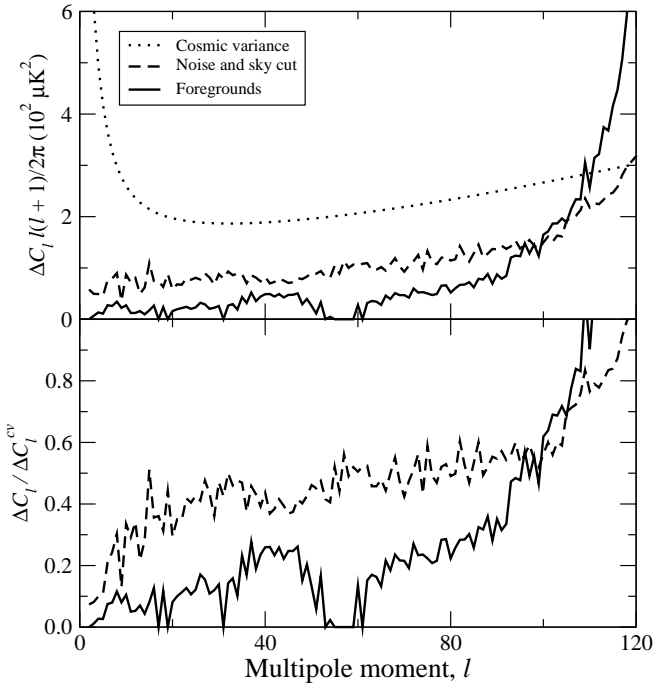}}
\caption{Top panel: Contributions to the total power spectrum
  uncertainty from cosmic variance (dotted line), sky cut and
  instrumental noise (dashed line) and foregrounds (solid line),
  respectively. Bottom panel: Ratio of noise and sky cut errors
  (dashed line) and foreground errors (solid line) to cosmic
  variance.}
\label{fig:error_comp}
\end{figure}

The top panel of Figure~\ref{fig:wmap_sigma} shows the posterior mean
realization specific spectrum, $\sigma_{\ell}$, for three different
cases. First, the true full-sky spectrum is plotted as a red line.
Second, the cut-sky but CMB-only spectrum is shown as blue curve, and
finally, the cut-sky and ``foreground-contaminated'' spectrum is shown
as a black curve. The $1\sigma$ confidence region about the latter is
marked as a gray region. The bottom panel shows the difference between
the foreground-contaminated spectrum and the full-sky spectrum (red),
and difference between the two cut-sky spectra (blue).

In terms of absolute differences, we see that the foreground errors
(blue curve in Figure \ref{fig:wmap_sigma}) are in general less than
$50\,\muK^2$ at $\ell \le 30$, with a few occasional peaks at
$\sim100\,\muK^2$, and without any striking biases. Already at this point,
we may thus predict that the absolute effect of residual temperature
foregrounds on cosmological parameters will be small at large angular
scales, when using the component separation method presented in this
paper.

\begin{figure*}[t]
\mbox{\epsfig{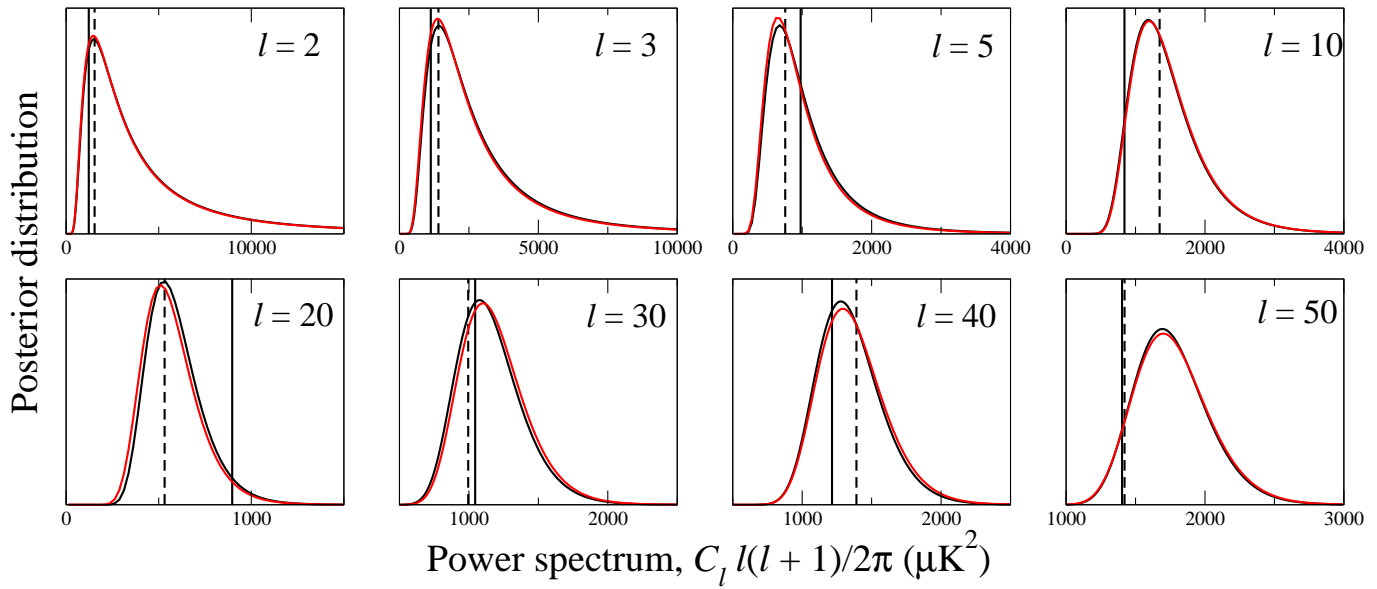}}
\caption{Slices through the CMB power spectrum likelihood. Black lines
  show the distributions from the foreground-free simulation, and red
  lines show the same for the simulation that did include foregrounds.
  The ensemble-averaged spectrum is indicated by vertical solid lines,
  and the true realization-specific spectrum by dashed lines. Note
  that at very low $\ell$'s, the distributions are essentially
  identical, because of cosmic variance domination. At $\ell \gtrsim
  30$--40, the additional uncertainty due to foregrounds start to
  become visible.}
\label{fig:posteriors}
\end{figure*}

Next, we consider the foreground induced uncertainties. First we note
that if the total CMB spectrum uncertainty has been properly
estimated, then the full-sky difference spectrum (red curve in 
Figure~\ref{fig:wmap_sigma}) should be distributed according to the
uncertainties indicated by the gray region. Except for some noticeable
correlated features around $\ell \approx 50$, this agreement is quite
reasonable. Second, the blue curve shows the differences due to
foregrounds alone. This term should thus be described by a
corresponding increase in the total uncertainty when including
foregrounds in the analysis.

To understand the relative magnitude of these contributions, it is
instructive to compute the relative magnitudes of the errors due to
cosmic variance, sky cut and instrumental noise, and foregrounds. 
These can be estimated from quantities ready at
hand. First, the standard expression for the cosmic variance is 
\begin{equation}
\textrm{Var}(\textrm{cosmic variance}) = \frac{2}{2\ell+1} C_{\ell}^2.
\end{equation}
Second, the uncertainty due to the mask and instrumental noise alone
is given by the variance of $\sigma_{\ell}^i$,
\begin{equation}
\textrm{Var}(\textrm{mask,noise}) = \langle(\sigma_{\ell}^i)^2 -
\langle\sigma_{\ell}^i\rangle^2\rangle,
\end{equation}
where $\sigma_{\ell}^{i}$ are generated in the analysis without
foregrounds.  Similarly, the uncertainty due to the combined effect of
the mask, instrumental noise, and foregrounds is given by the same
expression but computed from the samples that also include
foregrounds. To establish an order of magnitude approximation of the
foreground-induced uncertainty alone, we assume that the variances add
in quadrature,
\begin{equation}
  \textrm{Var}(\textrm{fg}) =
  \textrm{Var}(\textrm{mask,noise,fg}) - \textrm{Var}(\textrm{mask,noise}) 
\end{equation}
This is of course not strictly correct, because the errors in question
are quite non-Gaussian, but it is a sufficient approximation for our
purposes.

These three functions are shown in the top panel of Figure
\ref{fig:error_comp} for the data set described above. In the bottom
panel, we show the ratio of the mask and noise error and the
foreground error, respectively, to cosmic variance. First we note that
the foreground error is always smaller than the mask and noise induced
error, except at the very highest $\ell$'s, where the estimates are
anyway not reliable. However, at $\ell\sim40$ these two are almost
comparable in magnitude, both at the 25--50\% level of the cosmic
variance.  Once again assuming that these errors add in quadrature,
neglecting a 20\% error term implies underestimating the full errors
by about 2\% ($\sqrt{1+0.2^2}-1 = 0.02$).

At larger angular scales, we see that the foreground uncertainty
becomes essentially irrelevant, simply because the cosmic variance
dominates completely: At $\ell \le 30$, the relative magnitude of this
term is seldom more than 10\% of the cosmic variance, which translates
to a relative under-estimation of the total error by only 0.5\%. This
implies a somewhat surprising conclusion: Exact foreground error
propagation is not important on the very largest scales, below, say,
$\ell \le 30$, simply because the cosmic variance is so totally
dominating.  However, the same does not necessarily hold true on
smaller scales. At $\ell \approx 40$, the foreground error increases
the total uncertainty by several percent, and this is likely to
increase further to smaller scales\footnote{The peculiarly low
  foreground error at $\ell \approx 50$ may be connected with the lack
  of features in the dust template at these angular scales; we do not
  believe it is a general feature.}. It could constitute a very
significant fraction of the total error in the range between $\ell
\sim 50$--200, depending on the spatial foreground power spectrum.

In Figure \ref{fig:posteriors}, we plot eight slices through the power
spectrum likelihood between $\ell = 2$ and 50, for the two cases that
exclude (black curves) and include (red curves) foregrounds. In
these plots, we see the same behaviour as discussed above: At very
low $\ell$'s, the widths of the distributions with and without
foregrounds are essentially identical. However, at $\ell=40$ the
effect of the foreground uncertainties starts to become visible, as
the red distribution is noticeably wider than the black distribution.
Still, it is also very clear from this figure that the effect of
neglecting foreground errors in the temperature spectrum at $\ell
\lesssim 50$ in terms of cosmological parameters will be minimal.

To demonstrate this statement, we define a simple two-parameter power
spectrum model,
\begin{equation}
C_{\ell} = q \left(\frac{\ell}{15}\right)^n C_{\ell}^{\textrm{in}},
\label{eq:two_par_model}
\end{equation}
where $q$ is a free overall amplitude, $n$ is a spectral tilt
parameter, and $C_{\ell}^{\textrm{in}}$ is the actual theoretical
power spectrum used to generate the simulation.  We then map out the
corresponding two-dimensional posterior using the Blackwell-Rao
estimator \citep{chu:2005}. The analysis is restricted to the range
between $\ell=2$ and 30, which is the primary target range for our
first WMAP analysis \citep{eriksen:2007c}.

\begin{figure}[t]
\mbox{\epsfig{figure=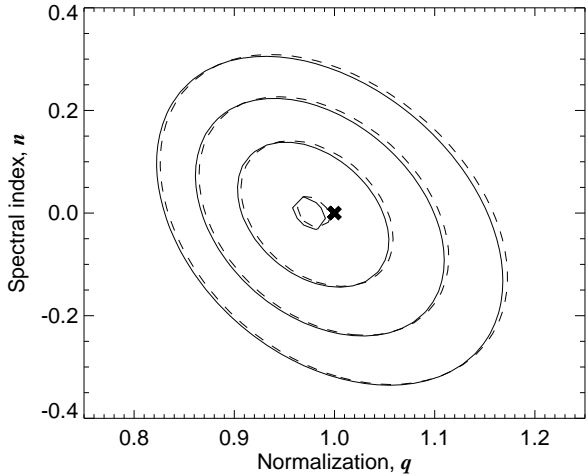,width=\linewidth,clip=}}
\caption{Joint posterior distributions for the two-parameter model
  defined by Equation \ref{eq:two_par_model}. Dashed contours show the
  results for the analysis without foregrounds, and solid contours
  show the same for the analysis with foregrounds. In both cases, the
  contours are where $-2\ln P(q,n|\Bd)$ rises by 0.1, 2.3, 6.17 and
  11.8 from its minimum value, corresponding (for Gaussian
  distributions) to the peak and the 1, 2, and $3\sigma$ confidence
  regions. The cross marks the true input value, $(q,n) = (1,0)$.}
\label{fig:qn_model}
\end{figure}

Figure~\ref{fig:qn_model} shows the results in terms of two sets of 
contours. The dashed contours show the results
from the analysis excluding foregrounds, and the solid contours show
the results including foregrounds. The true value of $(q,n)=(1,0)$ is
marked by a cross. The agreement between the two sets of results is
excellent.

Two conclusions may be drawn from this exercise. First, the method
presented in this paper is fully capable of extracting the valuable
cosmological signal from the 3-yr WMAP data at large angular scales in
quite realistic simulations, even when using the simplified foreground
model described earlier.  Second, the increased uncertainty in
cosmological parameters due to these foregrounds at $\ell \lesssim 30$
is negligible.

\section{Conclusions}
\label{sec:conclusions}

We have presented an algorithm for joint component separation and CMB
power spectrum estimation. This algorithm is a natural extension of
the CMB Gibbs sampler previously developed by \citet{jewell:2004},
\citet{wandelt:2004} and \citet{eriksen:2004b}, and the foreground
sampler described by \citet{eriksen:2006}. The basic product from this
algorithm is a set of joint samples drawn from the full joint
posterior $P(C_{\ell}, \mathbf{s}, \theta|\mathbf{d})$, where
$C_{\ell}$ is the CMB power spectrum, $\mathbf{s}$ is the CMB sky
signal, and $\theta$ denotes the set of all parameters in the
foreground model. With this tool, exact marginalization over very
general foreground models is feasible, and proper foreground
uncertainties may be propagated seamlessly through to the CMB power
spectrum and, therefore, to cosmological parameters.

There are some potential pitfalls the user of the method needs to be
aware of before applying it to real data. In particular, one has to
pay attention to possible degeneracies in the parametric signal model
fitted to the data. Such degeneracies are not uncommon in 
models relevant to CMB foreground analysis. Two specific examples are
synchrotron and free-free emission, with spectral indices of
$\beta_{\textrm{s}} \sim -3$ and $\beta_{\textrm{ff}} = -2.15$,
respectively, and the degeneracy between unknown offsets at all
frequency bands and the foreground zero-level. In order to obtain
reliable one-dimensional marginal estimates of each component
individually, one must either make sure that the data have sufficient
power to resolve the model, or impose priors to break the
degeneracies. Fortunately, since only the sum over all foregrounds is
relevant to the actual CMB reconstruction, these issues are of little
concern to the cosmological interpretation of the data.

The primary target of this work is Planck,  scheduled to be launched in
late 2008, and which will observe the microwave sky at nine frequencies from
30 to 857\,GHz. Combined with the five WMAP frequencies, these fourteen
sky maps will constitute an outstanding data set for both cosmological
and Galactic studies. Using the methods described in this paper, it
will be possible to constrain three or four foreground components
pixel-by-pixel, and even more if adopting spatial templates as priors
(e.g., the H$_{\alpha}$ template as a tracer for free-free).

A more immediate application is the analysis of the 3-yr WMAP
temperature data, which is presented in a separate Letter by
\citet{eriksen:2007c}. As demonstrated in the present paper, the
method is fully capable of extracting the valuable
cosmological signal at large angular scales from this data set, both
in terms of the CMB power spectrum and cosmological parameters.
Further, a very general foreground model may be constrained to within
a few percent in all parameters. 

In its present form, the method assumes identical beam responses at
each frequency band.  This limits its application to the lowest
angular resolution of a given data set. However, this is not a
fundamental limitation of the method, but only of our current
implementation. Specifically, it is straightforward to rewrite the
sampling equations presented in \S\ref{sec:hybrid_sampling} to
handle the foreground amplitudes in spherical harmonic space, similar
to the current treatment of the CMB sky signal. In that case,
convolution with individual beams at each frequency poses no problem.
An added bonus is that one may then optionally also either estimate or
impose a spatial power spectrum on the foregrounds, just as for the
CMB component. The implementation of this is left for future work.

Still, as demonstrated in this paper, even with the current
implementation we are able to perform a complete Bayesian joint CMB
and foreground analysis with the 3-yr WMAP temperature data at $\ell
\lesssim 30$--50. Simple extrapolation with respect to angular
resolution and instrumental noise suggests that we will be able to do
at least as well for Planck up to $\ell \sim 100$--200, beyond which 
other and simpler approaches may be applicable. We will quantify 
these projections in greater detail in an upcoming
publication where we apply
the method to state-of-the-art Planck-based simulations. An important
part of this work is to quantify the impact of modelling errors.

To conclude, given the recent successes of the Gibbs sampling approach
for analyzing both temperature and polarization CMB data, and now also
including realistic foreground modelling, we believe that
this method, or generalizations thereof, should be the baseline
analysis strategy for Planck at large angular scales. We also note
that its computational efficiency and unparallelled capabilities for
propagating systematic uncertainties is a combination that will
prove extremely valuable to future CMB experiments.

\begin{acknowledgements}
  We acknowledge use of the HEALPix software \citep{gorski:2005} and
  analysis package for deriving the results in this paper. We
  acknowledge use of the Legacy Archive for Microwave Background Data
  Analysis (LAMBDA). This work was partially performed at the Jet
  Propulsion Laboratory, California Institute of Technology, under a
  contract with the National Aeronautics and Space Administration. HKE
  acknowledges financial support from the Research Council of Norway, and 
  the hospitality of the Center for Long Wavelength Astrophysics at JPL..
\end{acknowledgements}

\appendix

\section{The template amplitude coupling matrix}
\label{app:coupling_matrix}

In Section \ref{sec:amplitude_sampling} we described the sampling
algorithm for the conditional Gaussian distribution $P(\Bs, a_{\nu,i},
b_j, \mathbf{c}_k|C_{\ell}, \theta_k, \Bd)$. This involved calculating
the joint mean,
\begin{equation}
\hat{{\bf x}} = \mathcal{A} \left[ \begin{array}{c}
\sum_{\nu} \BA^{t} \BN_{\nu}^{-1} \Bd \\
\mathbf{t}^{t}_{\nu,j} \BN_{\nu}^{-1} \Bd \\
\sum_{\nu} f_j(\nu) \mathbf{f}_j^{t} \BN^{-1} \Bd \\
\sum_{\nu} \mathbf{g}_k(\nu; \theta_k) \BN_{\nu}^{-1} \Bd
\end{array} \right],
\label{eq:joint_mean2}
\end{equation}
and multiplying with the inverse covariance matrix,
\begin{equation}
\mathcal{A}^{-1} = \left[ \begin{array}{cccc}
\BS^{-1} +  \BA^{t} \BN^{-1} \BA  &  \BA^{t} \BN^{-1} \BT & \BA^{t} \BN^{-1} \BF & \BA^{t}
\BN^{-1} \BG \\
\BT^{t} \BN^{-1} \BA & \BT^{t} \BN^{-1} \BT & \BT^{t} \BN^{-1} \BF & \BT^{t} \BN^{-1} \BG \\
\BF^{t} \BN^{-1} \BA & \BF^{t} \BN^{-1} \BT & \BF^{t} \BN^{-1} \BF & \BF^{t} \BN^{-1} \BG \\
\BG^t \BN^{-1} \BA & \BG^t \BN^{-1} \BT & \BG^t \BN^{-1} \BF & \BG^t \BN^{-1} \BG
\end{array} \right].
\label{eq:coupling_matrix2}
\end{equation}

We now write each element in this matrix explicitly, for the
benefit of readers who want to implement the algorithm themselves.
The elements in the first row of blocks are given by,
\begin{align}
\BS^{-1} +  \BA^{t} \BN^{-1} \BA  & \equiv  \BS^{-1} +
\sum_{\nu} \BA_{\nu}^{t} \BN_{\nu}^{-1} \BA_{\nu}\\
\BA^{t} \BN^{-1} \BT & \equiv  \BA_{\nu}^{t} \BN_{\nu}^{-1} \mathbf{t}_{\nu,j} \\
\BA^{t} \BN^{-1} \BF & \equiv  \sum_{\nu} \BA_{\nu}^{t} \BN_{\nu}^{-1} f_j(\nu)
\mathbf{f}_j \\
\BA^{t} \BN^{-1} \BG & \equiv  \sum_{\nu} \BA_{\nu}^{t} \BN_{\nu}^{-1}
\mathbf{g}_{k}(\nu; \theta_k)
\end{align}
The upper part of the second row is
\begin{align}
\BT^{t} \BN^{-1} \BT & \equiv  \mathbf{t}^t_{\nu,j} \BN^{-1} \mathbf{t}_{\nu,k}  \\
\BT^{t} \BN^{-1} \BF & \equiv  \sum_{\nu} \mathbf{t}^t_{\nu,j} \BN_{\nu}^{-1}
f_k(\nu) \mathbf{f}_j  \\
\BT^{t} \BN^{-1} \BG & \equiv  \sum_{\nu} \mathbf{t}^t_{\nu,j}
\BN_{\nu}^{-1} \mathbf{g}_{k}(\nu; \theta_k).
\end{align}
Note that in the above, $\BT^{T} \BN^{-1} \BT$ is the block containing
the second derivatives of the log density with respect to the
amplitudes $(a_{\nu,j}, a_{\nu', k})$, but by the assumed independence
of noise at different frequency channels, the terms with $\nu \neq
\nu'$ vanish. The elements of the upper part of the third row are
\begin{align}
\BF^{t} \BN^{-1} \BF & \equiv  \sum_{\nu}
f_j(\nu) \mathbf{f}^t_j \BN_{\nu}^{-1} \mathbf{f}_k f_k(\nu)  \\
\BF^{t} \BN^{-1} \BG & \equiv  
\sum_{\nu} f_j(\nu) \mathbf{f}^t_j
\BN^{-1}\mathbf{g}_{k}(\nu; \theta_k), 
\end{align}
and, finally, for the last block on the diagonal we have
\begin{equation}
\BG^t \BN^{-1} \BG = \sum_{\nu} \mathbf{g}^t_j(\nu; \theta_j)
\BN_{\nu}^{-1} \mathbf{g}_k(\nu; \theta_k).
\end{equation}

\section{Jeffreys' ignorance prior for a spectral index}
\label{sec:jeffreys_prior}

In \S\ref{sec:flat_vs_jeffrey} it was shown that to obtain
unbiased parameter estimates based on marginal statistics for a
power-law foreground model, it is necessary to adopt a proper
ignorance prior for the spectral index. In this Appendix we derive the
full expression for this prior, including noise and
the antenna-to-thermodynamic conversion factor, $a(\nu)$. 

The data model is in this case
\begin{equation}
d_{\nu} = A a(\nu) \left(\frac{\nu}{\nu_1}\right)^{\beta} + n_{\nu},
\label{eq:offset_model2}
\end{equation}
where $n_{\nu}$ is a noise term with variance $\sigma_{\nu}^2 =
\left<n_{\nu}^2\right>$. We assume no noise correlations between
frequencies. The log-likelihood then reads
\begin{equation}
-2  \ln \mathcal{L}(\beta) \propto \sum_{\nu} \left(\frac{d_{\nu} - A
      a(\nu)(\nu/\nu_1)^{\beta}}{\sigma_{\nu}^2}\right)^2.
\end{equation}

Jeffreys' ignorance prior is defined by 
\begin{equation}
  P_{\textrm{J}}(\beta) \sim \sqrt{F_{\beta\beta}} = \sqrt{-\left<\frac{\partial^2 \ln \mathcal{L}}
  {\partial^2\beta}\right>},
\end{equation}
where the averaging brackets denote an ensemble average. We therefore
differentiate the log-likelihood twice and find
\begin{equation}
  -2\frac{\partial^2 \ln \mathcal{L}}  {\partial^2\beta} = 4\sum_{\nu}
  \frac{1}{\sigma_{\nu}^2} \left\{A^2 a(\nu)^2
    \left(\frac{\nu}{\nu_1}\right)^{2\beta}
    \ln^2\left(\frac{\nu}{\nu_1}\right) -\left[d_{\nu} - A a(\nu)
      \left(\frac{\nu}{\nu_1}\right)\right] Aa(\nu)
    \left(\frac{\nu}{\nu_1}\right)^{\beta}
    \ln^2\left(\frac{\nu}{\nu_1}\right) \right\}
\end{equation}
Taking the ensemble average of this expression simply means setting
$\left<d_{\nu}\right> = A a(\nu)(\nu/\nu_1)^{\beta}$, and the second
term therefore vanishes,
\begin{equation}
  -\left<\frac{\partial^2 \ln \mathcal{L}}{\partial^2\beta}\right> \propto \sum_{\nu}
  \frac{A^2 a(\nu)^2}{\sigma_{\nu}^2}
  \left(\frac{\nu}{\nu_1}\right)^{2\beta} \ln^2\left(\frac{\nu}{\nu_1}\right),
\end{equation}
neglecting irrelevant constants. Thus, the full expression for
Jeffreys' ignorance prior for a spectral index $\beta$ reads
\begin{equation}
  P_{\textrm{J}}(\beta) \propto \sqrt{\sum_{\nu}
  \frac{a(\nu)^2}{\sigma_{\nu}^2}
  \left(\frac{\nu}{\nu_1}\right)^{2\beta}
  \ln^2\left(\frac{\nu}{\nu_1}\right)}
\label{eq:full_jeffreys_prior}
\end{equation}

\section{Template orthogonality constraints}
\label{sec:full_constraint}

Most of the discussion in \S\ref{sec:degeneracies} concerned the
degeneracy between free offsets at each frequency band and the
zero-level of an foreground component with a free amplitude at each
pixel. If neither is known, it is possible to add an arbitrary
constant to all amplitudes, and subtract a correspondingly scaled
value from each free offset, essentially without affecting the final
$\chi^2$.

To break this degeneracy, two approaches were proposed.  First, if
external calibration is possible at one or more frequencies, such
information should be exploited and conditioned upon. The second
approach was to demand that the free offsets do not have frequency
component similar to that of the foreground, by effectively fitting
out the corresponding spectrum from the offsets.

These issues apply more generally to any fixed template with a free
amplitude at all frequency bands. For most typical applications, this
includes also three unknown dipole components, in addition to the
familiar offset or monopole. In the following, we simply consider a
general collection of templates denoted by $\BT$, which is an $\npix
\times N_{\textrm{t}}$ matrix consisting of $N_{\textrm{t}}$ templates
listed in its columns. Coupled to this, we define a coefficient vector
$\Ba_{\nu} = \{a_{\nu,i}\}$ containing the template amplitudes for each
frequency and template. The sky response at frequency $\nu$ is thus
given by $\BT \Ba_{\nu}$.

We now derive the joint orthogonality constraints for $N_{\textrm{t}}$
templates, for a sky model that includes two components with a free
amplitude at each pixel, namely a CMB sky signal and a foreground
component with a given frequency spectrum.  We denote the vectors of
free template coefficients for the CMB and foreground terms by $\Bb$
and $\Bc$, respectively, and the foreground spectrum is defined by the
$\npix \times \npix$ diagonal matrix $\BF_{\nu}$ having entries equal
to $g(\nu;\theta)$ on the diagonal. Note that the frequency spectrum
of the CMB component is constant, and the corresponding matrix is
therefore the identity. It is omitted in the following.

With this notation, the $\chi^2$ to be minimized reads
\begin{equation}
\chi^2 = \sum_{\nu} \left(\Ba_{\nu}^t \BT^t - \Bb^t \BT^t - \Bc^t
  \BF_{\nu}^t \BT^t\right) \BN_{\nu}^{-1} \left(\BT \Ba - \BT \Bb -
  \BT \BF_{\nu} \Bc\right)
\end{equation}
Equating the derivatives of this expression with respect to $\Bb$ and
$\Bc$ to zero gives
\begin{align}
\frac{\partial \chi^2}{\partial\Bb} &= -2\sum_{\nu} \BT^t
\BN_{\nu}^{-1} \left(\BT \Ba - \BT \Bb -  \BT \BF_{\nu} \Bc\right) = 0\\
\frac{\partial \chi^2}{\partial\Bc} &= -2\sum_{\nu} \BF_{\nu}^t\BT^t
\BN_{\nu}^{-1} \left(\BT \Ba - \BT \Bb -  \BT \BF_{\nu} \Bc\right) = 0
\end{align}
These two sets of equations provide a general expression for $\Ba$ as
a function of $\Bb$ and $\Bc$, and at this point we have done nothing
more than performed a partial change of basis.

We now impose the prior that breaks the degeneracy between the
template and the foreground amplitudes, by requiring $\Bc = 0$. The
two above equations then has a unique solution. In particular, $\Ba$
is given by
\begin{equation}
\sum_{\nu} (\BB_{\nu} - \BD\BC^{-1}\BA_{\nu})\Ba = 0,
\label{eq:joint_constraints}
\end{equation}
where we, for notational transparency, have defined four ancillary
matrices
\begin{align}
\BA_{\nu} &= \BT^t \BN_{\nu}^{-1} \BT \\
\BB_{\nu} &= \BF^t\BT^t \BN_{\nu}^{-1} \BT \\
\BC &= \sum_{\nu} \BA_{\nu} \\
\BD &= \sum_{\nu} \BB_{\nu}.
\end{align}
Note that Equation \ref{eq:joint_constraints} corresponds to
$N_{\textrm{t}}$ separate constraints on $\Ba$, and, in particular,
each row of the matrix in parentheses defines one orthogonality vector
$\mathbf{q}_{\nu}$. These constraints are thus imposed in the CG
solver using the same projection operator method that was described in
\S\ref{sec:amplitude_sampling}.

However, we once again stress that this approach corresponds to
imposing a very strong prior that is not realized in practice: If
there are indeed random fluctuations in the unknown offsets, then
these will have a component that happens to have a spectrum similar to
the foregrounds. This component will in the present approach be
interpreted as a foreground signal. Further, the sampler is by this
constraint not allowed to explore the joint distribution between the
two components. In other words, both the marginal zero-level and
offset uncertainties will be under-estimated by this approach.  (Note
that most other parameters, such as the CMB signal, are very weakly
affected by this, because they only depend on the sum of the two
components, not each of the two separately.)

That being said, for experiments where this constraint is required
(e.g., differential observatories such as WMAP), it is difficult
indeed to construct an alternative and more self-consistent approach.
For example, the WMAP team adopted a method based on a co-secant fit
to a very crude, plane parallel galaxy model. In their case, no
uncertainties were quoted at all. This specific issue is considered in
further detail in the actual 3-yr WMAP analysis \citep{eriksen:2007c}.



\begin{thebibliography}{}

\bibitem[Ade et al.(2007)]{ade:2007} 
Ade, P., et al.\ 2007, ApJ, submitted, [arXiv:0705.2359]

\bibitem[Baccigalupi et al.(2001)]{baccigalupi:2001} 
Baccigalupi, C., Burigana, C., Perrotta, F., De Zotti, G., La Porta,
L., Maino, D., Maris, M., \& Paladini, R.\ 2001, \aap, 372, 8

\bibitem[Baccigalupi(2003)]{baccigalupi:2003} 
Baccigalupi, C.\ 2003, New Astronomy Review, 47, 1127

\bibitem[Baccigalupi et al.(2004)]{baccigalupi:2004} 
Baccigalupi, C., Perrotta, F., de Zotti, G., Smoot, G.~F., Burigana,
C., Maino, D., Bedini, L., \& Salerno, E.\ 2004, \mnras, 354, 55

\bibitem[Banday et al.(1996)]{banday:1996} Banday, A.~J., G{\'o}rski,
  K.~M., Bennett, C.~L., Hinshaw, G., Kogut, A., \& Smoot, G.~F.\
  1996, \apjl, 468, L85

\bibitem[Barreiro et al.(2004)]{barreiro:2004} 
Barreiro, R.~B., Hobson, M.~P., Banday, A.~J., Lasenby, A.~N.,
Stolyarov, V., Vielva, P., \& G{\' o}rski, K.~M.\ 2004, \mnras, 351,
515

\bibitem[Bennett et al.(2003a)]{bennett:2003a} 
Bennett, C.~L.~et al.~2003a, \apjs, 148, 1

\bibitem[Bennett et al.(2003b)]{bennett:2003b} Bennett, C.~L., et al.\ 
2003b, \apjs, 148, 97 

\bibitem[Bond et al.(1998)]{bond:1998} Bond, J.~R., Jaffe, A.~H., 
\& Knox, L.\ 1998, \prd, 57, 2117 

\bibitem[Bouchet \& Gispert(1999)]{bouchet:1999} 
Bouchet, F.~R., \& Gispert, R.\ 1999, New Astronomy, 4, 443

\bibitem[Box \& Tiao(1992)]{box:1992}
Box, G. E. P, \& Tiao, G. C. 1992, \emph{Bayesian Inference in
  Statistical Analysis}, Wiley Classics Library edition,
John Wiley and Sons

\bibitem[Brandt et al.(1994)]{brandt:1994} 
Brandt, W.~N., Lawrence, C.~R., Readhead, A.~C.~S., Pakianathan,
J.~N., \& Fiola, T.~M.\ 1994, \apj, 424, 1

\bibitem[Chu et al.(2005)]{chu:2005} Chu, M., Eriksen, H.~K., Knox,
  L., G{\'o}rski, K.~M., Jewell, J.~B., Larson, D.~L., O'Dwyer, I.~J.,
  \& Wandelt, B.~D.\ 2005, \prd, 71, 103002

\bibitem[Delabrouille et al.(2003)]{delabrouille:2003} 
Delabrouille, J., Cardoso, J.-F., \& Patanchon, G.\ 2003, \mnras, 346,
1089

\bibitem[Dickinson et al.(2003)]{dickinson:2003} Dickinson, C.,
  Davies, R.~D., \& Davis, R.~J.\ 2003, \mnras, 341, 369

\bibitem[Donzelli et al.(2006)]{donzelli:2006} Donzelli, S., et al.\ 
2006, \mnras, 369, 441 

\bibitem[Efstathiou(2004)]{efstathiou:2004} Efstathiou, G.\ 2004, 
\mnras, 349, 603 

\bibitem[Eriksen et al.(2004a)]{eriksen:2004a} Eriksen, H.~K., Banday, 
A.~J., G{\'o}rski, K.~M., \& Lilje, P.~B.\ 2004a, \apj, 612, 633 

\bibitem[Eriksen et al.(2004b)]{eriksen:2004b} 
Eriksen, H.~K., et al.\ 2004b, \apjs, 155, 227

\bibitem[Eriksen et al.(2006)]{eriksen:2006} 
Eriksen, H.~K., et al.\ 2006, \apj, 641, 665

\bibitem[Eriksen et al.(2007a)]{eriksen:2007a} Eriksen, H.~K., et al.\
2007a, \apj, 656, 641

\bibitem[Eriksen et al.(2007b)]{eriksen:2007b} Eriksen, H.~K., Huey, 
G., Banday, A.~J., G{\'o}rski, K.~M., Jewell, J.~B., O'Dwyer, I.~J., \& 
Wandelt, B.~D.\ 2007b, \apjl, 665, L1 

\bibitem[Eriksen et al.(2007c)]{eriksen:2007c} Eriksen, H.~K.,
  Dickinson, C., Jewell, J.~B., Banday, A.~J., G{\'o}rski, K.~M., \&
  Lawrence, C. R. 2007c, \apjl, submitted, [arXiv:0709.1037]

\bibitem[Finkbeiner et al.(1999)]{finkbeiner:1999} 
Finkbeiner D.P., Davis M., \& Schlegel D.J. 1999, ApJ, 524, 867

\bibitem[Finkbeiner(2003)]{finkbeiner:2003} 
Finkbeiner, D.~P.\ 2003, \apjs, 146, 407

\bibitem[Gelfand \& Smith(1990)]{gelfand:1990}
Gelfand, A. E., \& Smith, A. F. M. 1990, J. Am. Stat. Asso., 85, 398

\bibitem[Gelman \& Rubin(1992)]{gelman:1992}
Gelman, A., \& Rubin, D. 1992, Stat. Sci., 7, 457

\bibitem[Giardino et al.(2002)]{giardino:2002} 
Giardino, G., Banday, A.~J., G{\' o}rski, K.~M., Bennett, K., Jonas,
J.~L., \& Tauber, J.\ 2002, \aap, 387, 82

\bibitem[G{\'o}rski(1994)]{gorski:1994} G{\'o}rski, K.~M.\ 1994, \apjl, 
430, L85

\bibitem[G{\'o}rski et al.(1996)]{gorski:1996} G{\'o}rski, K.~M.,
  Banday, A.~J., Bennett, C.~L., Hinshaw, G., Kogut, A., Smoot, G.~F.,
  \& Wright, E.~L.\ 1996, \apjl, 464, L11

\bibitem[G{\'o}rski(1997)]{gorski:1997} G{\'o}rski, K.~M.\ 1997, Proc.
  XVIth Moriond Astrophysics Meeting, Microwave Background
  Anisotropies, Editions Frontieres, Gif-sur-Yvette, p.77,
  [astro-ph/9701191]

\bibitem[G{\'o}rski et al.(2005)]{gorski:2005} 
  G{\' o}rski, K.~M., Hivon, E., Banday, A.~J., Wandelt, B.~D.,
  Hansen, F.\,K., Reinecke, M., \& Bartelmann, M. 2005, \apj, 622, 759

\bibitem[Haslam et al.(1982)]{haslam:1982} 
Haslam, C. G. T., Salter, C. J., Stoffel, H., \& Wilson, W.\ 1982,
\aaps, 47, 1

\bibitem[Hinshaw et al.(2007)]{hinshaw:2007} Hinshaw, G., et al.\ 
2007, \apjs, 170, 288 

\bibitem[Hivon et al.(2002)]{hivon:2002} 
Hivon, E., G{\' o}rski, K.~M., Netterfield, C.~B., Crill,
  B.~P.,Prunet, S., \& Hansen, F.\ 2002, \apj, 567, 2

\bibitem[Hobson et al.(1998)]{hobson:1998} 
Hobson, M.~P., Jones, A.~W., Lasenby, A.~N., \& Bouchet, F.~R.\
1998,\mnras, 300, 1

\bibitem[Jeffreys(1961)]{jeffreys:1961}
Jeffreys, H. 1961, \emph{Theory of probability}, third edition,
Oxford, Clarendon Press

\bibitem[Jewell et al.(2004)]{jewell:2004} 
  Jewell, J., Levin, S., \& Anderson, C.  H.  2004, \apj, 609, 1

\bibitem[Kuo et al.(2007)]{kuo:2007} Kuo, C.~L., et al.\ 2007, 
\apj, 664, 687 

\bibitem[Larson et al.(2007)]{larson:2007} Larson, D.~L., Eriksen,
H.~K., Wandelt, B.~D., G{\'o}rski, K.~M., Huey, G., Jewell, J.~B., \&
O'Dwyer, I.~J.\ 2007, \apj, 656, 653

\bibitem[Lewis \& Bridle(2002)]{lewis:2002} 
  Lewis, A., \& Bridle, S.\ 2002, \prd, 66, 103511

\bibitem[Liu(2001)]{liu:2001}
Liu, J. S. 2001, \emph{Monte Carlo Strategies in Scientific Computing},
Springer

\bibitem[Maino et al.(2002)]{maino:2002} 
Maino, D., et al.~2002, \mnras, 334, 53

\bibitem[Maino et al.(2003)]{maino:2003} 
Maino, D., Banday, A.~J., Baccigalupi, C., Perrotta, F., \&
  G{\'o}rski, K.~M.\ 2003, \mnras, 344, 544

\bibitem[Montroy et al.(2006)]{montroy:2006} Montroy, T.~E., et al.\ 
2006, \apj, 647, 813

\bibitem[O'Dwyer et al.(2004)]{odwyer:2004} O'Dwyer, I.~J., et al.\
2004, \apjl, 617, L99

\bibitem[Page et al.(2007)]{page:2007} Page, L., et al.\ 2007, 
\apjs, 170, 335 

\bibitem[Readhead et al.(2004)]{readhead:2004} Readhead, A.~C.~S., et 
al.\ 2004, \apj, 609, 498

\bibitem[Shewchuk(1994)]{shewchuk:1994} Shewchuk, J.  R.  1994,
  http://www.cs.cmu.edu/$\sim$quake-papers/painless-conjugate-gradient.ps 

\bibitem[Sievers et al.(2007)]{sievers:2007} Sievers, J.~L., et al.\ 
2007, \apj, 660, 976 

\bibitem[Smith et al.(2007)]{smith:2007} Smith, K.~M., Zahn, O., 
\& Dor{\'e}, O.\ 2007, \prd, 76, 043510 

\bibitem[Spergel et al.(2007)]{spergel:2007} Spergel, D.~N., et al.\ 
2007, \apjs, 170, 377 
 
\bibitem[Stivoli et al.(2006)]{stivoli:2006} Stivoli, F., 
Baccigalupi, C., Maino, D., \& Stompor, R.\ 2006, \mnras, 372, 615 

\bibitem[Stolyarov et al.(2002)]{stolyarov:2002} 
Stolyarov, V., Hobson, M.~P., Ashdown, M.~A.~J., \& Lasenby, A.~N.\
  2002, \mnras, 336, 97

\bibitem[Stolyarov et al.(2005)]{stolyarov:2005} 
Stolyarov, V., Hobson, M.~P., Lasenby, A.~N., \& Barreiro, R.~B.\
  2005, \mnras, 357, 145

\bibitem[Szapudi et al.(2001)]{szapudi:2001} Szapudi, I., Prunet, 
S., \& Colombi, S.\ 2001, \apjl, 561, L11 

\bibitem[Tegmark \& Efstathiou(1996)]{tegmark:1996} 
Tegmark, M., \& Efstathiou, G.\ 1996, \mnras, 281, 1297

\bibitem[Tegmark et al.(2003)]{tegmark:2003} 
Tegmark, M., de Oliveira-Costa, A., \& Hamilton, A.~J.\ 2003, \prd,
68, 123523

\bibitem[Wandelt et al.(2004)]{wandelt:2004} 
  Wandelt, B.~D., Larson, D.~L., \& Lakshminarayanan, A.\ 2004, \prd,
  70, 083511

\bibitem[Wright et al.(1994)]{wright:1994} Wright, E.~L., Smoot, 
G.~F., Bennett, C.~L., \& Lubin, P.~M.\ 1994, \apj, 436, 443 


\end{thebibliography}
\end{document}